\documentclass[a4paper,fleqn,usenatbib]{mnras}

\usepackage{newtxtext,newtxmath}

\usepackage[T1]{fontenc}
\usepackage{ae,aecompl}

\usepackage{graphicx}  
\usepackage{amsmath}   
\usepackage{amssymb}   

\newcommand{\gccm}{\,g cm$^{-3}$}

\newcommand{\kms}{\,km s$^{-1}$}
\newcommand{\CO}{\,C$^{18}$O}

\newcommand{\half}{\frac{1}{2}}

\title[Accretion driven turbulence in filaments]{Accretion driven turbulence in filaments I: Non-gravitational accretion}

\author[S. Heigl et al.]{
S. Heigl,$^{1,2}$\thanks{E-mail: heigl@usm.lmu.de}
A. Burkert,$^{1,2}$
M. Gritschneder,$^{1}$
\\
$^{1}$Universit\"ats-Sternwarte, Ludwig-Maximilians-Universit\"at M\"unchen, Scheinerstr. 1, 81679 Munich, Germany\\
$^{2}$Max-Planck Institute for Extraterrestrial Physics, Giessenbachstr. 1, 85748 Garching, Germany
}

\date{Accepted XXX. Received YYY; in original form ZZZ}

\pubyear{2017}

\begin{document}
\label{firstpage}
\pagerange{\pageref{firstpage}--\pageref{lastpage}}
\maketitle

\begin{abstract}
  We study accretion driven turbulence for different inflow velocities
  in star forming filaments using the code \textsc{ramses}. Filaments
  are rarely isolated objects and their gravitational potential will lead to
  radially dominated accretion. In the non-gravitational case, accretion by
  itself can already provoke non-isotropic, radially dominated turbulent motions
  responsible for the complex structure and non-thermal line widths observed in
  filaments. We find that there is a direct linear relation between the absolute
  value of the total density weighted velocity dispersion and the infall
  velocity. The turbulent velocity dispersion in the filaments is independent of
  sound speed or any net flow along the filament. We show that the density weighted
  velocity dispersion acts as an additional pressure term supporting the
  filament in hydrostatic equilibrium. Comparing to observations, we find that the
  projected non-thermal line width variation is generally subsonic independent
  of inflow velocity.
\end{abstract}




\section{Introduction}
\label{sec:introduction}

   Turbulent motions are ubiquitous on all astrophysical scales. There is evidence
   for highly complex non-thermal motion from the intergalactic medium to the
   interstellar medium (ISM) and individual molecular clouds down to even the smallest
   scales of protostellar discs. Likewise, as part of the ISM, star forming filaments
   are no exception to this observational fact. In contrast to its importance the origin
   of turbulence is still not fully understood and there are numerous potential sources
   for turbulent motions in the ISM
   \citep{maclow2004,elmegreen2004,elmegreen2010,klessen2014}. On cloud scale,
   molecular line observations are dominated by supersonic motions and show a direct
   correlation between size and line width \citep{larson1981}. This is usually
   interpreted as the direct result of a turbulent cascade from the scale of
   a tens of parsec sized molecular cloud down to the scale of parsec sized filaments
   \citep{kritsuk2013,federrath2016,padoan2016}.
   Filaments then inherit their internal velocity dispersion from the motions on
   larger scales. This model is also favoured by \textit{Herschel} observations
   which show that filamentary structures are ubiquitous in molecular clouds
   \citep{andre2010,arzoumanian2011,arzoumanian2013,andre2014}. This picture has
   recently been challenged by the discovery that more massive filaments are
   actually complex bundles of fibers whose line-of-sight superposition create the
   observed supersonic linewidths \citep{hacar2013} and which were also found
   to form in numerical simulations \citep{smith2014,moeckel2015}. Independent of the
   formation process, we argue that accretion driven by the gravitational potential of
   the filaments alone can be enough to stimulate turbulent motions that are in
   agreement with the observations (see also \citealp{ibanez2016}).

   Recent images of the high column density gas, traced by the usually optically thin
   \CO\ line of nearby filaments have shown that the non-thermal linewidth is
   predominantly sub- and transsonic along filaments. This is true for L1517 in Taurus
   \citep{hacar2011}, where the mean is about half the sound speed, as well as for the
   fibre-like substructure of the L1495/B213 region in Taurus \citep{hacar2013}, where
   the mean is about the sound speed. Moreover, even in the Musca filament, a 6 pc long
   structure, subsonic non-thermal linewidths dominate along the filament
   \citep{hacar2016}. Larson was the first to connect the velocity dispersion
   to the size of molecular clouds \citep{larson1981}. While not strictly valid
   for filamentary structures, it predicts a factor of about three times higher
   velocity dispersion corresponding to the supersonic regime for a filament like
   Musca with a width of 0.14 pc \citep{cox2016}. Although still within the spread of
   the overall relation, Musca is definitely an outlier in terms of its length and it
   is unclear which process leads to a structure of that size and that low velocity
   dispersion. In this paper we explore a possible origin of turbulence in filaments.

   In the following sections, we introduce the basic concepts we use to
   constrain our model (\autoref{sec:concepts}). We then discuss the code and the
   numerical set-up (\autoref{sec:numericalsetup}). Thereafter, we present our
   results of the simulations and discuss them in detail (\autoref{sec:simulations}).
   Additionally, we show that turbulence plays a role in creating a hydrostatic
   equilibrium (\autoref{sec:pequilibrium}). Finally, we compare our data to the
   observations (\autoref{sec:observations}) and investigate the dependence on filament
   inclination.

\section{Basic concepts}
\label{sec:concepts}

   In order to sustain turbulence inside a filament there has to be an external driving
   mechanism. Otherwise, turbulent motions decay on the timescale of a crossing time
   \citep{maclow1998,stone1998,padoan1999,maclow1999,maclow2004}. Here, we discuss a
   possible source of the external driving and the theoretical prediction.

   \subsection{Gravitational accretion onto a filament}

   Although there are different ways to accrete mass onto a filament, e.g. a converging
   flow, these processes are typically limited in time. A counterexample for a radial
   converging flow which is stable over longer timescales is the gravitational
   attraction of the filament itself. If we assume that the filament is
   isothermal and in hydrostatic equilibrium, then it has a density profile first
   described by \citet{stodolkiewicz1963} and \citet{ostriker1964}:
   \begin{equation}
     \rho(r) = \frac{\rho_c}{\left(1+\left(r/H\right)^{2}\right)^{2}}
   \end{equation}
   where $r$ is the cylindrical radius and $\rho_c$ is its central density. The
   radial scale height $H$ is given by
   \begin{equation}
      H^2 = \frac{2c_s^2}{\pi G \rho_c}
   \end{equation}
   where $c_s$ is the isothermal sound speed and $G$ the gravitational
   constant. We assume that the gas has an isothermal temperature of
   $10 \text{ K}$. Using a molecular weight of $\mu=2.36$ gives the
   isothermal sound speed of $c_s=0.19$ \kms. One can integrate
   the profile to $r \rightarrow \infty$ to get the critical line mass of
   \begin{equation}
      \left(\frac{M}{L}\right)_\text{crit} =
      \frac{2c_s^2}{G}\approx1.06\cdot 10^{16}
      \text{g cm}^{-1}\approx16.4 \text{ M}_{\sun}\text{ pc}^{-1}
   \end{equation}
   above which a filament will collapse under its self-gravity. Following
   \citet{heitsch2009}, for a given line mass $M/L$ the gravitational
   acceleration of the filament is:
   \begin{equation}
      a = -\frac{2GM/L}{r}
   \end{equation}
   One can calculate the potential energy which a gas parcel of mass $m$ loses in
   free-fall starting with zero velocity at a distance $R_0$ to the filament
   radius $R$ by integrating over $r$:
   \begin{equation}
     E_\text{pot} = 2 G (M/L) m \ln\left(\frac{R_0}{R}\right)
   \end{equation}
   Therefore, the inflow velocity at the point of accretion $R$ is:
   \begin{equation}
     v_r = 2\sqrt{G(M/L) \ln\left(\frac{R_0}{R}\right)}
   \end{equation}
   Note that similar to the free fall velocity, the value does not depend on the mass
   of the gas parcel. It is also not sensitive to neither the starting position nor
   the line mass while depending stronger on the latter. As the filament accretes mass
   and increases in line-mass, the inflow velocity grows. However, as we want to
   analyse accretion driven turbulence in an equilibrium state we keep the
   inflow velocity constant. Therefore we choose to neglect the effects of gravity
   and use an artificial but constant mass inflow where we set the inflow velocity
   also to a constant value. The effects of gravity will be discussed in a subsequent
   paper. Assuming the extreme case of a filament with a gravitational influence of a
   hundred times the filament radius, which for a typical radius of 0.05 pc
   \citep{arzoumanian2011} is the size of a typical molecular cloud, we still need a
   line mass which is several times higher than the critical line-mass to achieve an
   inflow velocity of even Mach 10.0. Thus, we limit our maximum inflow velocity to
   Mach 10.0.

   A consequence of a constant inflow velocity is a constant mass accretion rate. The
   absolute value is set by the radius of the inflow region $R_0$ and the density
   at that radius $\rho_0$:
   \begin{equation}
     \dot{M} = \rho_0 v_r 2 \pi R_0 L
     \label{eq:mdot}
   \end{equation}
   This should stay constant for every radial shell and thus leads to the following
   density profile outside of the filament:
   \begin{equation}
     \rho(r) = \rho_0 \frac{R_0}{r}
     \label{eq:rho}
   \end{equation}
   As there is an isothermal accretion shock formed at the filament boundary, pressure
   equilibrium requires that the mean density inside the filament is the outside density
   times a factor of the Mach number $\mathcal{M}$ squared. This leads to the following
   filament mass-radius relation:
   \begin{equation}
     M(R) = \rho_0 \mathcal{M}^2 \pi R R_0 L
     \label{eq:mass}
   \end{equation}
   This has to be the same as the accreted mass given by the mass accretion rate
   given by \autoref{eq:mdot} times the time $t$. Therefore the radius of the
   filament evolves as:
   \begin{equation}
     R(t) = \frac{2c_s^2 t}{v_r}
     \label{eq:rad}
   \end{equation}

   \subsection{Turbulence driven by accretion}

   Following \citet{klessen2010}, \citet{heitsch2013} derived an analytical
   expression for the velocity dispersion depending on the inflow velocity.
   We expect turbulence to decay on the timescale of a crossing time:
   \begin{equation}
     \tau_d \approx \frac{L_d}{\sigma}
   \end{equation}
   where $\sigma$ is the velocity dispersion in three dimensions and $L_d$ is the
   driving scale of the system. \citet{klessen2010} use the approach that the change
   in turbulent kinetic energy is given by the balance of the accretion of kinetic
   energy and the dissipation of turbulent energy:
   \begin{equation}
     \dot{E}_t = \dot{E}_{a} - \dot{E}_d = (1-\epsilon)\dot{E}_{a}
   \end{equation}
   with the energy accretion rate
   \begin{equation}
     \dot{E}_{a} = \half \dot{M} v_r^2
   \end{equation}
   and the loss by dissipation as
   \begin{equation}
     \dot{E}_d \approx \frac{E_{t}}{\tau_d}= \half \frac{M\sigma^3}{L_d}.
     \label{eq:ediss}
   \end{equation}
   They also introduce an efficiency factor $\epsilon$ as fraction of accreted energy
   which can sustain the turbulent motions:
   \begin{equation}
     \epsilon = \left|\frac{\dot{E}_d}{\dot{E}_a}\right|
   \end{equation}
   Thus, if the driving scale is the filament diameter $L_d = 2R$, \citet{heitsch2013}
   predicted a turbulent velocity dispersion of
   \begin{equation}
     \sigma = \left(2\epsilon R(t) v_r^2\frac{\dot{M}}{M(t)}\right)^{1/3}
     \label{eq:eequ}
   \end{equation}
   This assumes that $\epsilon$ is independent of the inflow velocity. In our simple
   case the radius depends on the constant inflow velocity as $1/v_r$ and the radial
   accretion leads to a radius and mass growing linear in time. Therefore, we
   expect a constant level of velocity dispersion which should behave as
   \begin{equation}
     \sigma \sim v_r^{1/3}.
   \end{equation}
   This is the relationship we want to confirm or disprove using numerical methods.

\section{Numerical set-up}
\label{sec:numericalsetup}

   We executed the numerical simulations with the code \textsc{ramses}
   \citep{teyssier2002}. The code uses a second-order Godunov scheme
   to solve the conservative form of the discretised Euler equations
   on an cartesian grid. For the simulations we applied the MUSCL
   scheme (Monotonic Upstream-Centred Scheme for Conservation Laws,
   \citet{vanLeer1977}) together with the HLLC-Solver \citep{toro1994}
   and the multidimensional MC slope limiter \citep{vanLeer1979}.

\begin{figure}
      \includegraphics[width=\columnwidth]{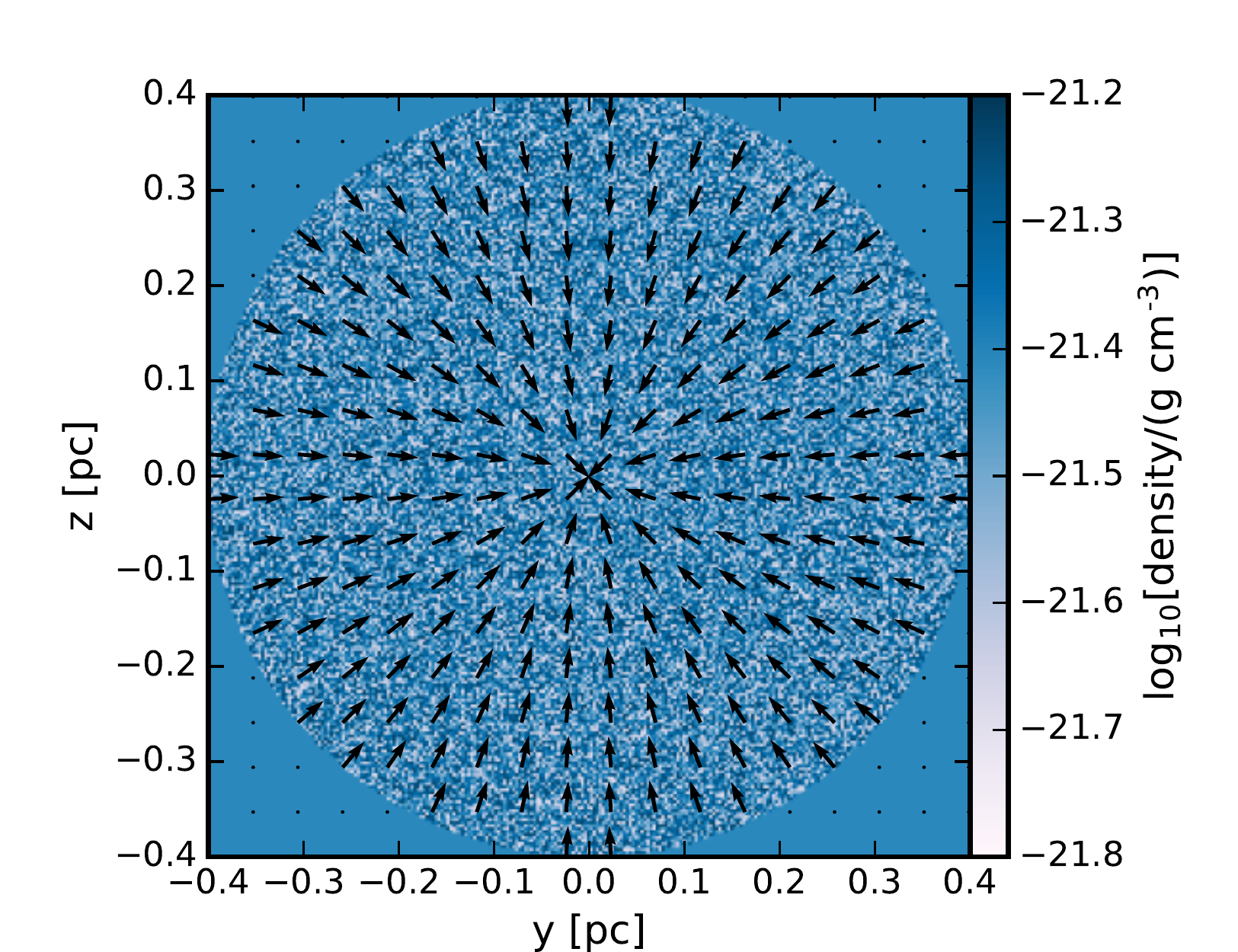}
      \caption{Density cut through the y-z plane of the initial conditions for all
      simulations. The material inside the inflow zone is perturbed with a random
      perturbation and has a constant velocity directed to the central line of the
      box as shown by the black arrows.}
      \label{fig:inicondition}
   \end{figure}

   We simulate a converging radial flow onto a non-gravitating filament in
   order to study the generated turbulence. We use a 3D box with periodic
   boundary conditions in the x-direction and outflow boundaries in the other
   directions. The periodic boundary prohibits the loss of turbulent motions in
   x-direction. As \textsc{ramses} cannot use a radial inflow boundary we define a
   cylindrical inflow zone which lies at the edges of the box and has a thickness of
   two cells from where material flows onto the central x-axis of the box. The initial
   gas density inside of the box is set to a mean of $3.92\cdot10^{-21}$\gccm,
   corresponding to about $10^2$ particles per cubic centimeters for a molecular weight
   of $\mu=2.36$. Additionally, a random perturbation is added
   inside of the inflow zone at the beginning of the simulation. This is illustrated in
   \autoref{fig:inicondition} where we show a density slice through the y-z plane.
   The inflow has a constant density of $3.92\cdot10^{-21}$\gccm and a constant
   velocity. Thus, it leads to a build-up of material in form of a filament
   with a radius that grows over time as it is not restricted by gravity. Consequently,
   one has to ensure a big enough box that an equilibrium can be established.
   The surrounding cells around the inflow zone are given a constant density with the
   same value as the inflow zone and pressure and do not affect the simulation.

   The complete box is set to be isothermal with a temperature of 10 K and
   a molecular weight of $\mu=2.36$. In general, the boxsize is 0.8 pc.
   For the control runs with a higher temperature the boxsize is doubled to
   ensure enough space to reach a velocity dispersion equilibrium.

   As the inflow initially leads to a thin, compressed central filament with a
   high density, we employ adaptive mesh refinement (AMR) to resolve high density
   regions while keeping the resolution low in the remainder of the box. We
   test different refinement strategies by varying the maximum resolution to fulfill the
   Truelove criterion for the maximum occurring density \citep{truelove1997} while
   keeping the minimum resolution constant at $256^3$ cells. Our initial approach is to
   resolve the Jeans length by 16 cells. We also test a more conservative criterion
   suggested by \citet{federrath2011} using 32 cells to resolve the Jeans length in
   order to sufficiently resolve the turbulent cascade. However, we cannot detect a
   quantitative difference in the velocity dispersion to the previous case with lower
   maximum resolution. Furthermore, even a lower maximum resolution of 8 cells per
   Jeans length does not show any difference in the value of the velocity dispersion.
   Despite there being no change in behavior for the maximum resolved density, we see a
   difference for a varying minimum resolution. We present the details and a resolution
   study in the next section.

\section{Simulations}
\label{sec:simulations}

   \begin{figure}
      \includegraphics[width=\columnwidth]{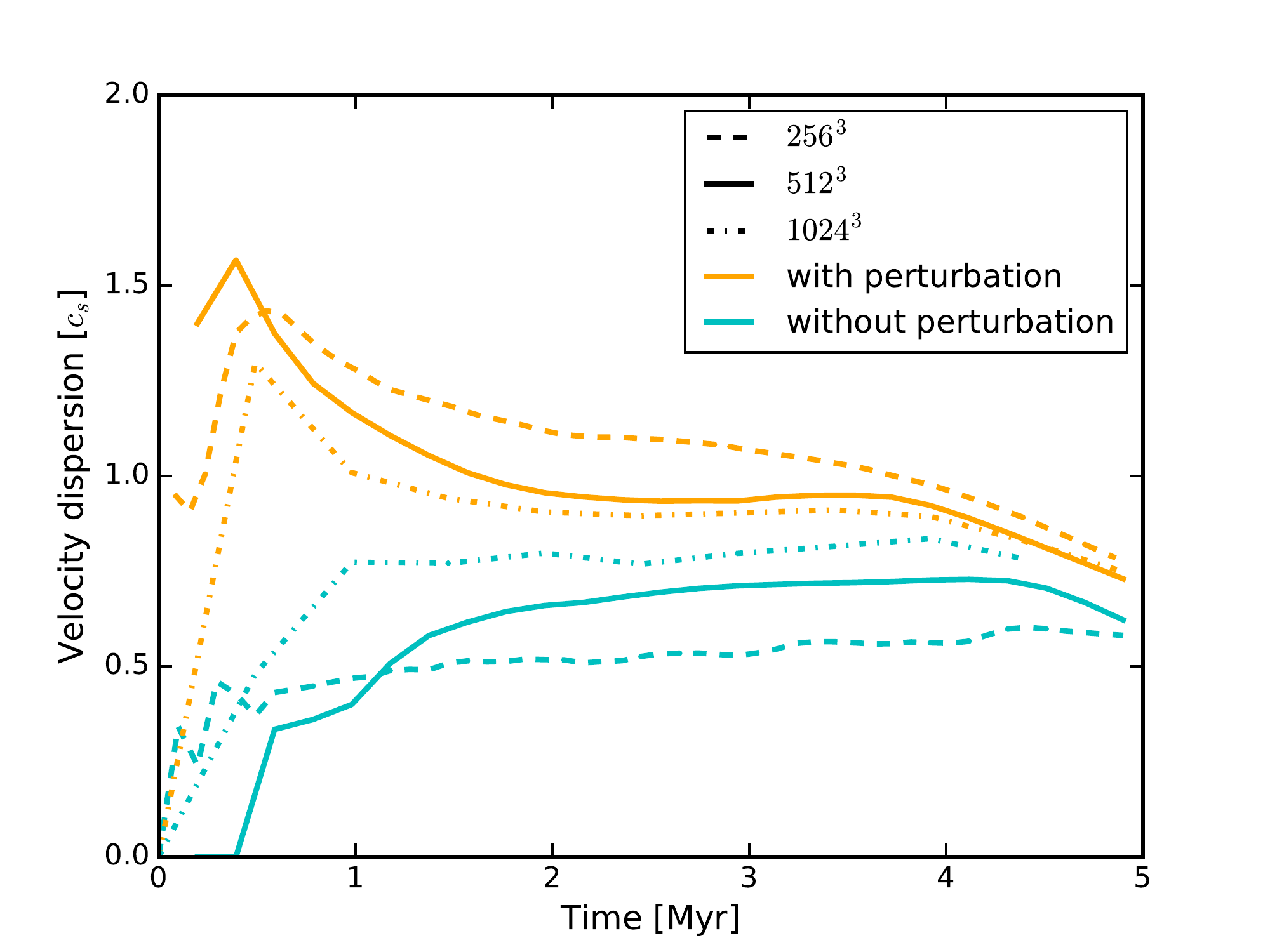}
      \caption{Evolution of the density weighted total velocity
      dispersion for the reference case with and without an initial density
      perturbation and for varying minimum resolution. After a short settling phase
      an equilibrium is established where the value of the velocity dispersion is
      constant. The simulations which do not include an initial density perturbation
      converge to the same equilibrium level as the ones including an initial density
      perturbation. Our later analysis is based on the case represented by the solid
      lines.}
      \label{fig:sigzero}
   \end{figure}

   \begin{figure*}
     \includegraphics[width=2.0\columnwidth]{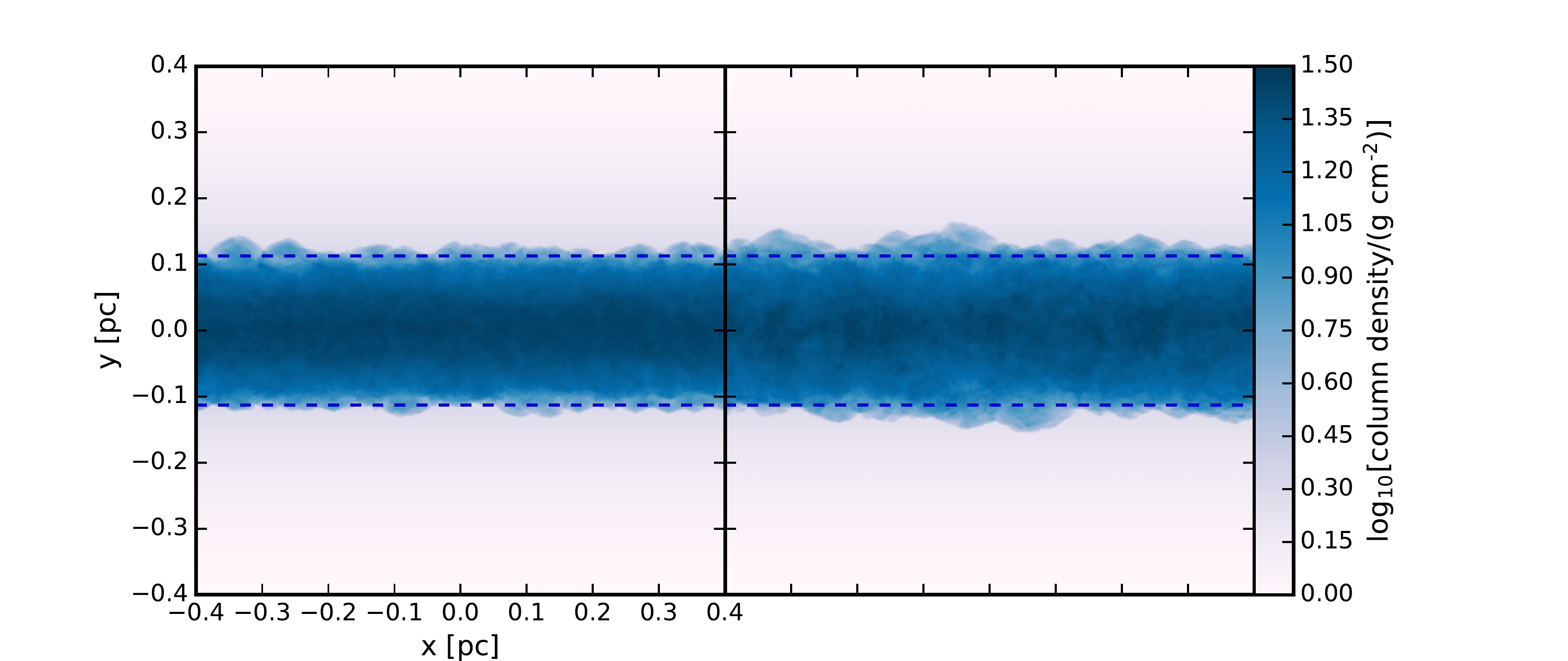}
      \caption{Projection of the highest minimum resolution test cases of a
      filament without initial perturbation on the left compared to one including an
      initial perturbation on the right after 1.5 Myr. Both cases have an inflow of
      Mach 5.0. The horizontal dashed lines indicate
      the analytical prediction for the radius.}
      \label{fig:slicezero}
   \end{figure*}

   \begin{figure}
      \includegraphics[width=\columnwidth]{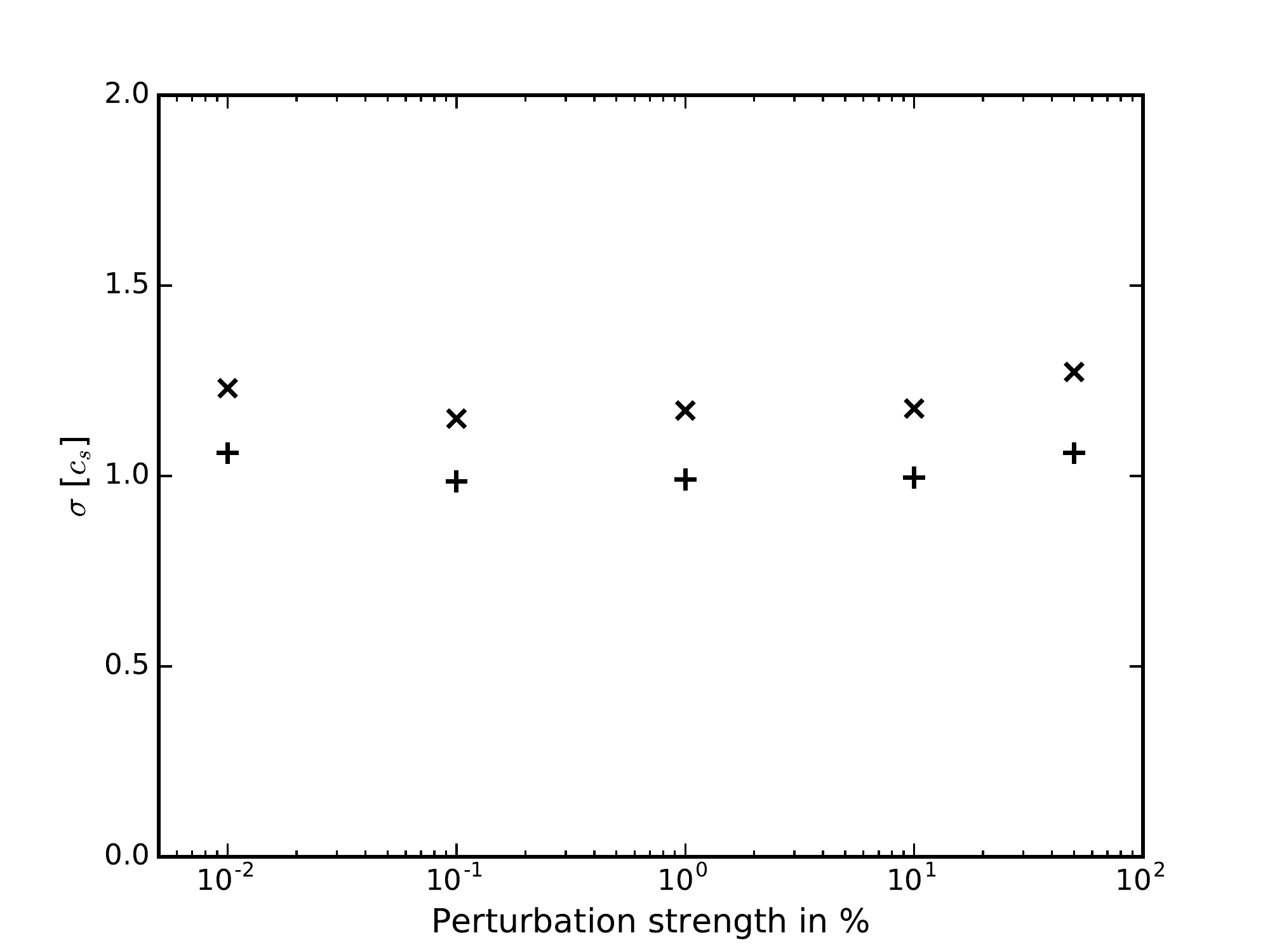}
      \caption{Value of the generated velocity dispersion with varying perturbation
      strength for the same resolution. The points show simulations including
      an initial density perturbation. The crosses give the volume weighted and the
      pluses the density weighted velocity dispersion.}
      \label{fig:pert}
   \end{figure}

   In this section we present the outcome of our simulations.
   In order to measure the velocity dispersion and the mass of the filament one has
   to distinguish between filament material and ambient medium. As the
   inflowing material is shocked at the filament surface there is a clear increase
   in density and a clear drop in radial velocity. As the internal density of the
   filament decreases over time due to \autoref{eq:mass}, we use the radial
   velocity to distinguish inflowing material from filament material instead of a
   density threshold. To measure the filament radius we use the mean
   position of the highest density gradient which traces the general shock position.

   \subsection{Initial perturbation}
   \label{sec:ini}

   \begin{figure}
      \includegraphics[width=\columnwidth]{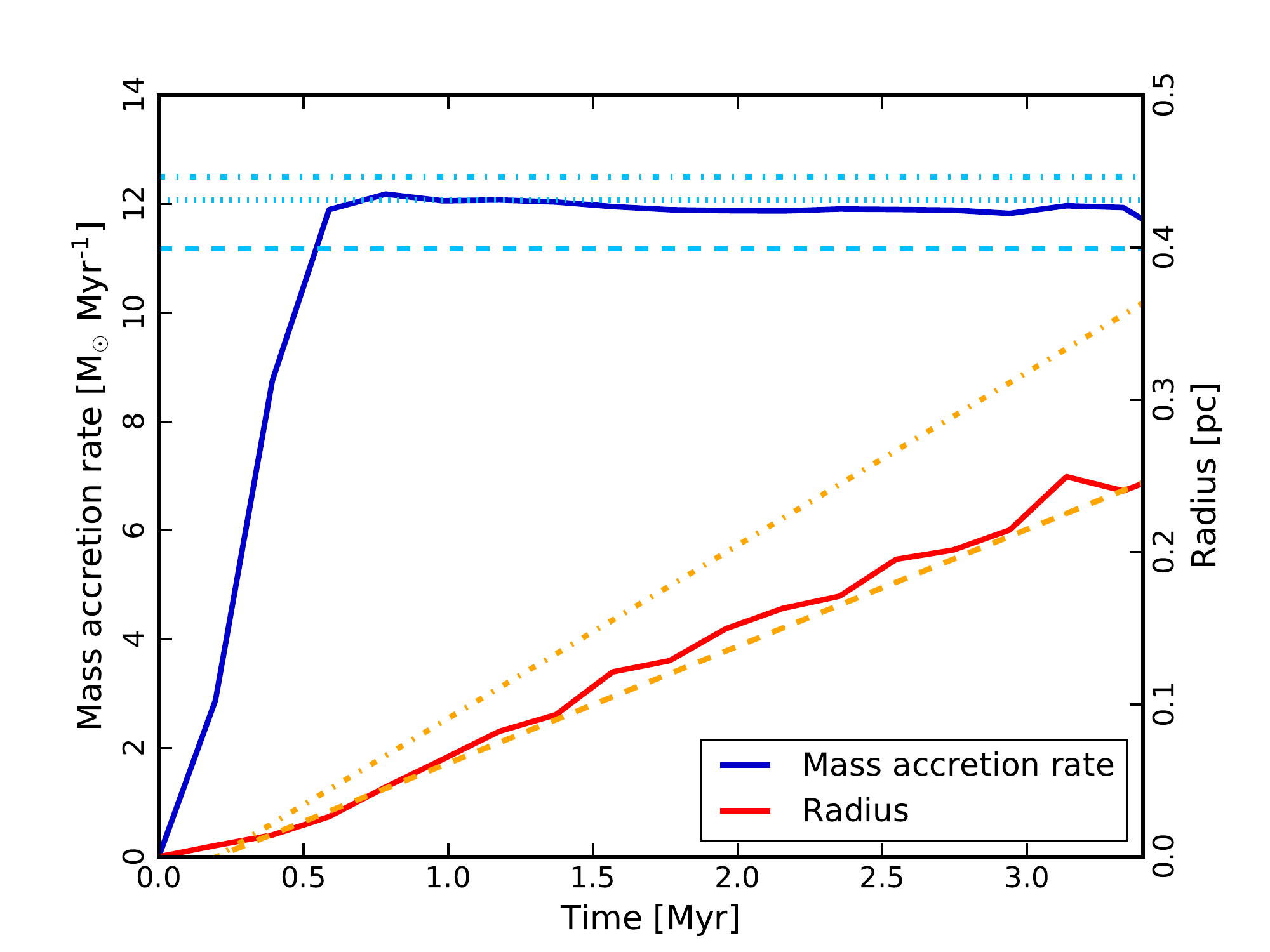}
      \caption{Evolution of the mass accretion rate (blue) and the filament radius
      (red) for the reference case with an initial density perturbation. The
      analytical predictions are given by the light blue and orange dashed lines
      respectively. The mass accretion rate is higher than the prediction but
      matches it if we take the radial expansion of the filament into account,
      as shown by the dotted line. Correcting both terms with a radial turbulent
      pressure overpredicts the measured evolution as can be seen by the
      dashed-dotted lines.}
      \label{fig:dtzero}
   \end{figure}

   As a reference test case we set up a converging flow with a velocity of Mach 5.0
   with and without a perturbation of the initial density field.
   As material streams in, it is compressed on the central axis of the box.
   In order to avoid unphysical densities in the initial phase of the case
   without initial perturbation we add an already existing
   filament to the initial conditions with a radius of 0.05 times the boxlength and
   with a constant density of ten times the ambient density corresponding to
   a total mass of $0.23$ M$_\odot$. Note that we do not need to include an initial
   filament for the simulations including an initial
   perturbation, as the generated turbulence prevents the density from reaching values
   which cannot be resolved.
   For a low resolution there is a stark contrast in the structure
   between both cases. An initial perturbation leads to a considerable amount of
   substructure which resembles observed filaments. The case lacking an initial
   perturbation does not obviously exhibit signs of turbulence being present. The
   motions are purely angular and radial and the little
   substructure which is generated is obscured due to the projection.
   However, as the inflow is perfectly symmetrical, only radial motions
   should exist. In order to understand this effect we vary the minimum resolution
   of the simulation and calculate the velocity dispersion over time.
   The volume weighted velocity dispersion is defined as the standard deviation of the
   spatial velocity distribution. We calculate the total velocity dispersion
   by taking the square root of the sum of the variances of the spatial
   components. For the density weighted velocity
   dispersion we normalise the velocities with the density of its respective cell
   divided by the average density before calculating the variance. Note that we use the
   central axis of the box to define the centre of the filament. As we use a
   cartesian grid code this can lead to a wrong split-up in the cylindrical
   components if the filament axis lies not exactly in the centre of the box. We
   analysed the error in the total velocity dispersion using a cartesian and a
   cylindrical calculation and the difference is at most one per cent
   and therefore negligible.
   We show the evolution of the density weighted velocity dispersion in
   \autoref{fig:sigzero}. One can see that an equilibrium is established after
   about two Myr where the velocity dispersion becomes almost constant. We observe
   this behavior in all our simulations, with and without initial perturbation,
   and it is also found in turbulent smooth-particle-hydrodynamical simulations of
   filaments forming in a turbulent medium \citep{clarke2017}.
   Varying the minimum resolution, one can see that the equilibrium level of
   the simulations including a perturbation decreases for higher resolution showing no
   more substantial change going from $512^3$ to $1024^3$ in minimal resolution. In
   contrast, the simulations without an initial perturbation develop higher and higher
   values of turbulence the better the resolution and begin to converge to a similar
   value as the cases with an initial perturbation. Moreover, even the visual
   impression of the turbulent structure for the cases with and without initial
   perturbations becomes increasingly similar with higher resolution. This is shown in
   \autoref{fig:slicezero} where we compare the cases of highest minimal resolution.
   We interpret this in such a way that there is a inherent physical level of
   generated turbulence from accretion. As long as there is a source of perturbation
   the growth in turbulence will tend to this value. In the case of a smooth initial
   inflow the perturbation is given by numerical noise which is smeared out for lower
   minimal resolutions due to numerical viscosity. For higher resolution, numerical
   viscosity is low enough that the numerical noise can grow to the
   same level of turbulence as in the case containing an added perturbation. The
   interstellar medium is not completely smooth and will always contain density
   fluctuations. In order to validate if the generated velocity dispersion depends on
   the initial density perturbation we vary the initial density
   perturbation in strength and form for a constant minimum resolution. Using a flat,
   "white noise" perturbation on the one hand and a Gaussian perturbation on the other
   shows no quantitative difference in magnitude of the intrinsic filament velocity
   dispersion. Varying the perturbation strength also shows no dependence on
   the initial perturbation. In \autoref{fig:pert} we show that even changing the
   perturbation amplitude over five orders of magnitude, the resulting value of velocity
   dispersion varies only minimally. As there is no more significant change for a
   minimum resolution of $1024^3$ all following analysis is carried out for a minimum
   resolution of $512^3$.

   \begin{figure*}
      \includegraphics[width=1.75\columnwidth]{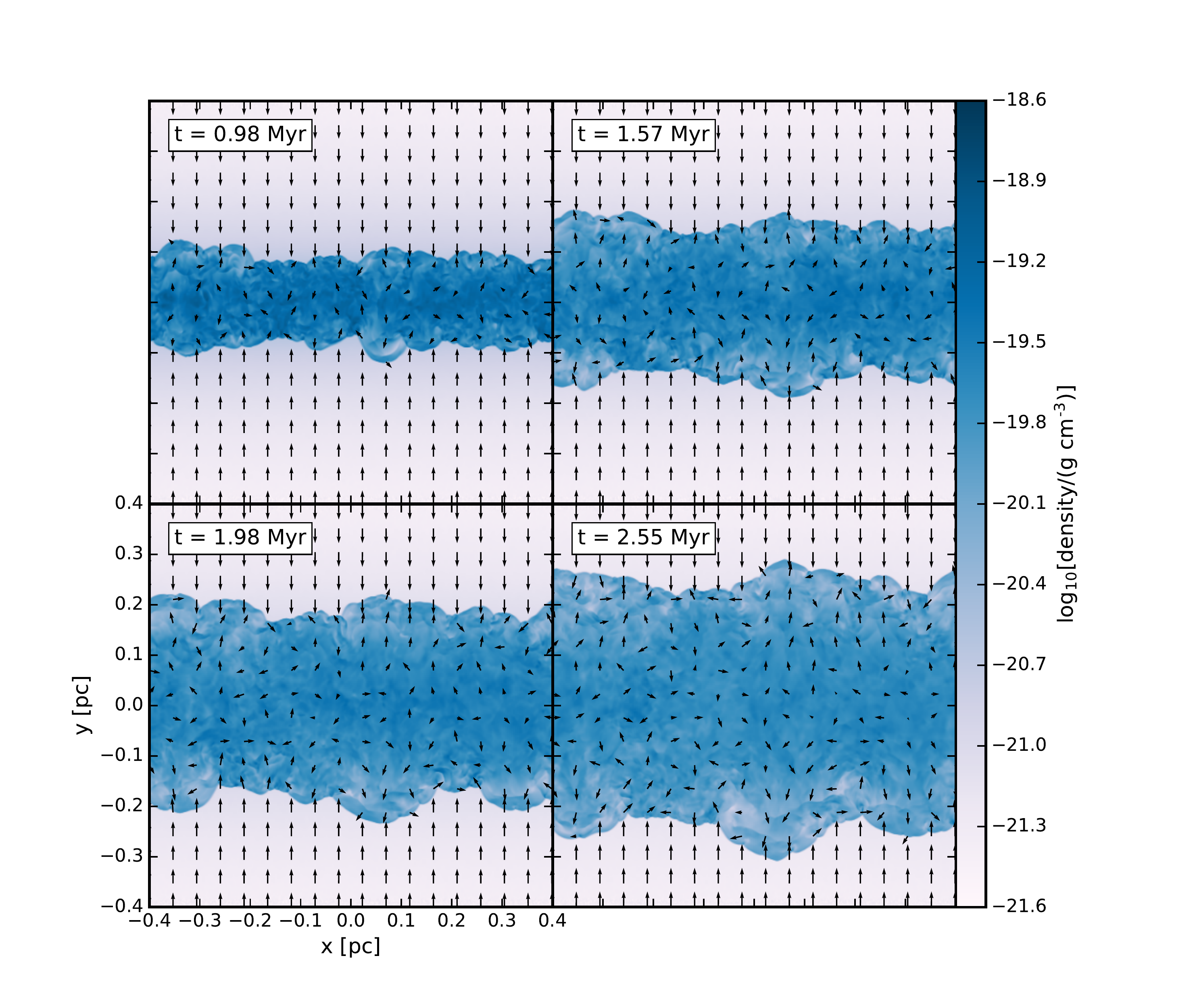}
      \caption{Density cut through the centre of the filament with a Mach 5.0 inflow.
      The arrows show the log-scaled velocities in the plane. In the inflow
      region their length corresponds to Mach 5.0 and they are normalised to zero at
      Mach 0.01. The filament shows clear signs of ongoing turbulent motions.}
      \label{fig:evosuper}
   \end{figure*}

   \begin{figure}
      \includegraphics[width=\columnwidth]{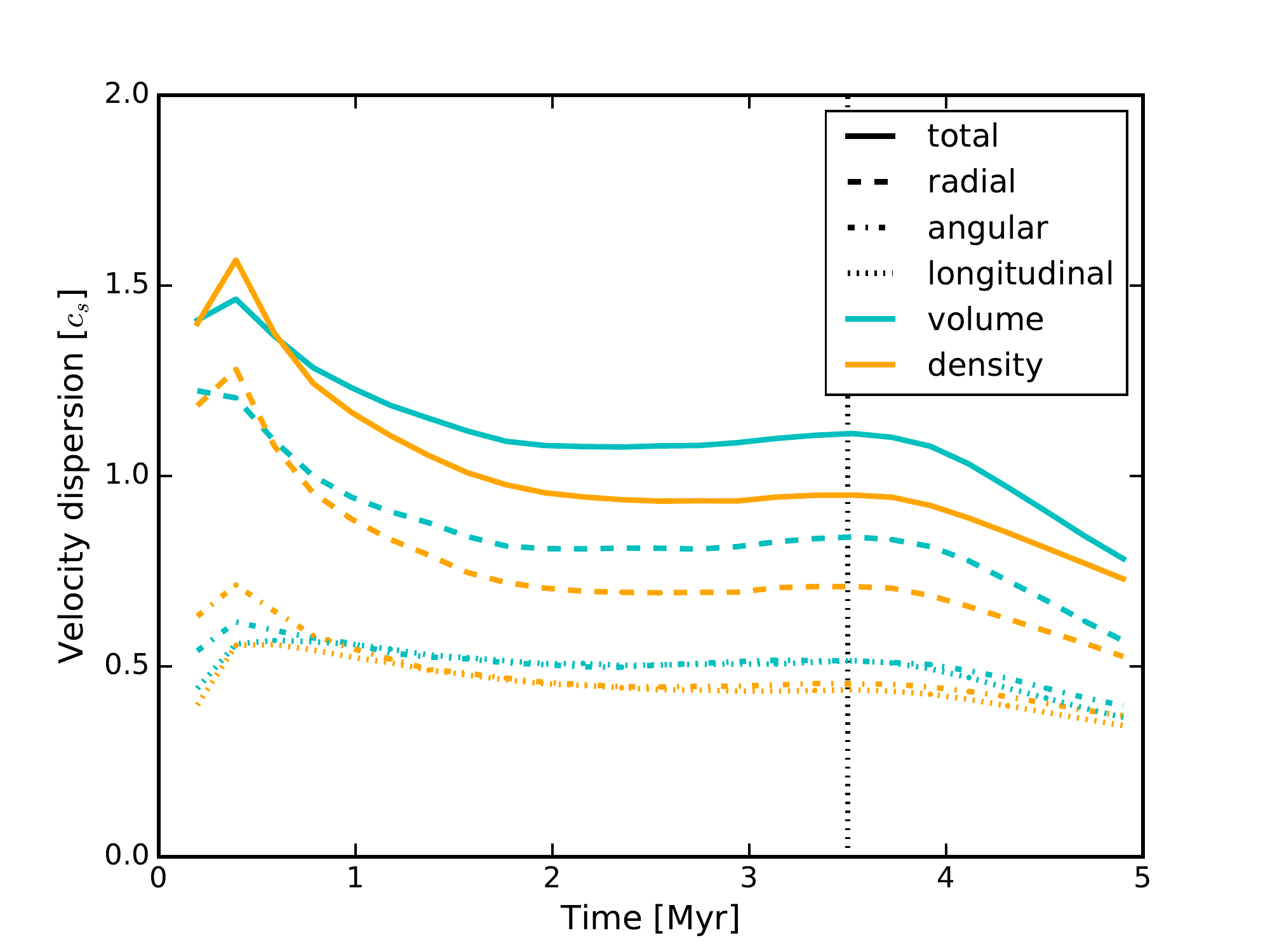}
      \caption{Evolution of the velocity dispersion of a Mach 5.0 inflow
      with an initial density perturbation. The volume weighted values are given
      by the cyan lines, the density weighted values by the orange lines. As in
      the reference case, an equilibrium is established after two Myr. The dotted
      vertical line shows the timestep when the filament radius reaches the domain
      boundary.}
      \label{fig:sigsuper}
   \end{figure}

   \begin{figure}
      \includegraphics[width=\columnwidth]{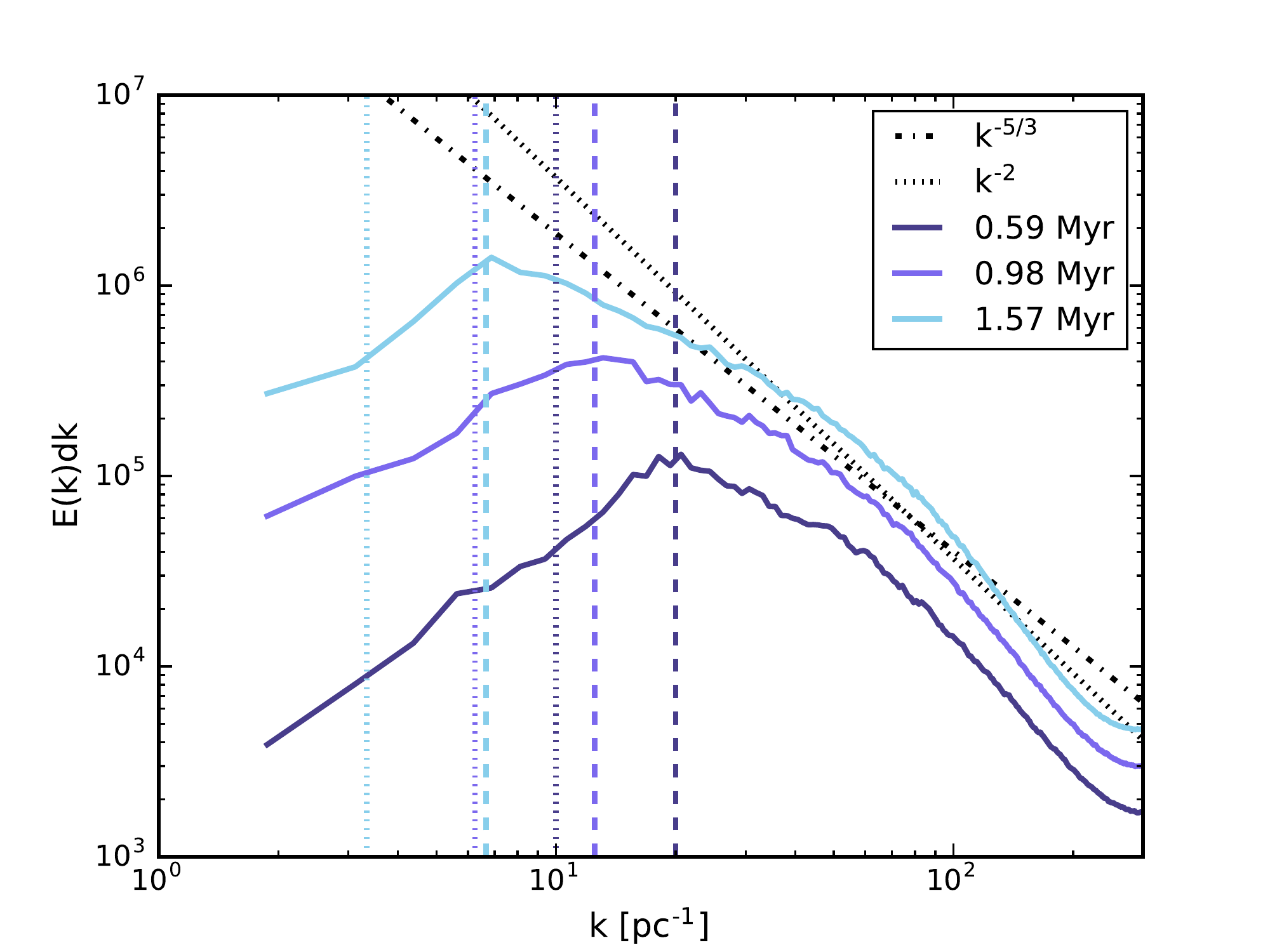}
      \caption{The kinetic energy power spectrum of the filament with Mach 5.0 inflow
      at different times. The black dashed-dotted line shows the expectation from
      Kolmogorov's theory and the black dotted line the expectation of Burgers
      turbulence. The vertical dashed lines show the respective filament radius and
      the vertical dotted lines the filament diameter.}
      \label{fig:superps}
   \end{figure}

   \begin{figure}
      \includegraphics[width=\columnwidth]{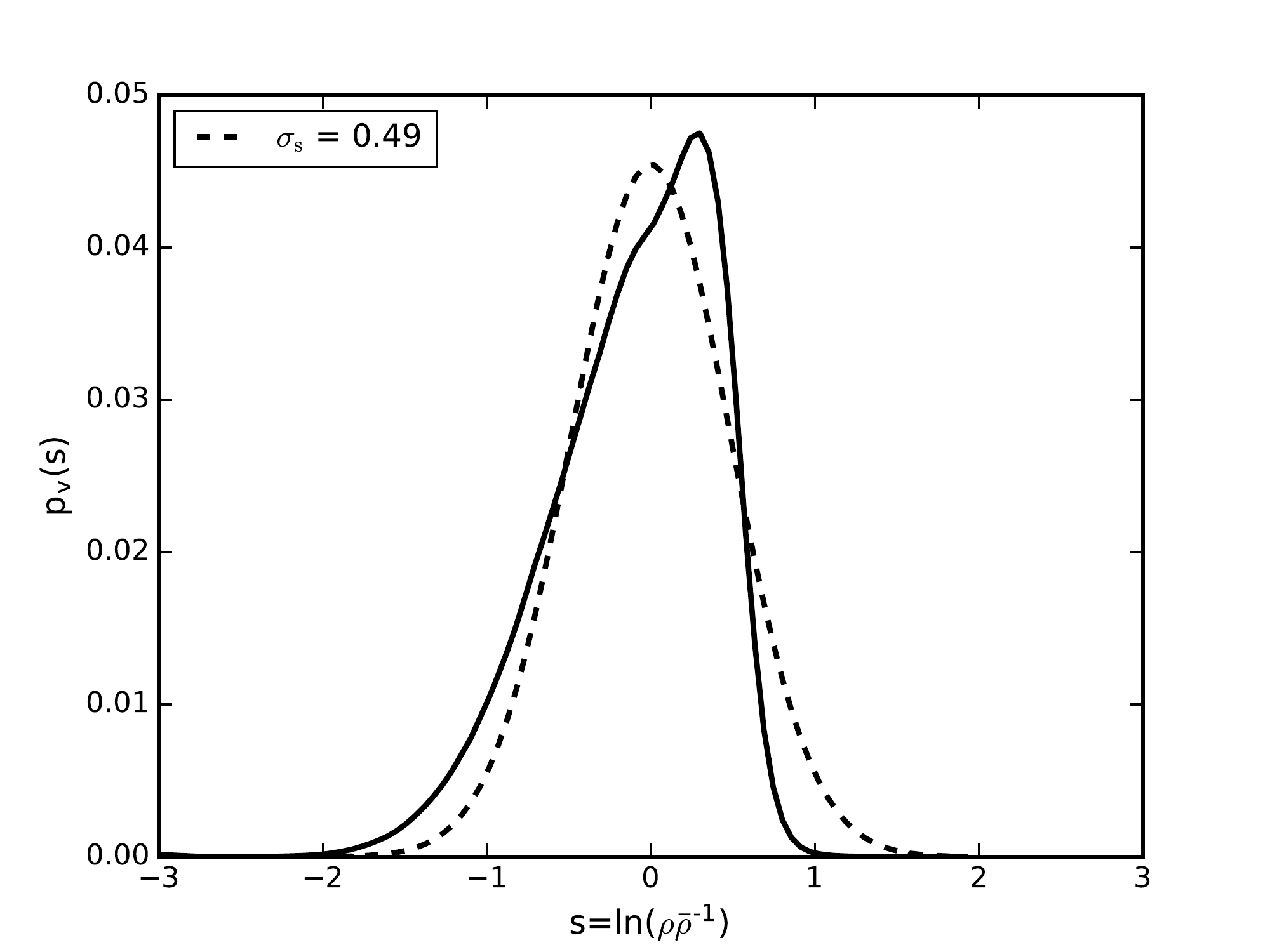}
      \caption{Volume weighted PDF of the logarithmic density
      $s=\ln(\rho \bar\rho^{-1})$. The density follows largely a log-normal
      distribution. The Gaussian fit leads to a standard deviation of 0.56.}
      \label{fig:dpdf}
   \end{figure}

   We also evaluate the radial evolution and mass accretion rate of the filament
   including an initial perturbation in order to verify if \autoref{eq:mdot}
   and \autoref{eq:rad} hold. In \autoref{fig:dtzero} we show both of them together
   with the analytical expectations. As our analytical prediction, the measured
   mass accretion rate is constant. Despite being close to the predicted value given
   by the blue dashed line, it has a small but measurable constant
   offset. This effect can be explained if we add the increase in mass
   accretion given by the expansion of the filament. Adding the radial growth velocity
   to \autoref{eq:mdot} gives the blue dotted line. This effect itself also should lead
   to a faster radial growth but the effect is minimal as we cannot see an significant
   offset of the analytical expectation for filament radius $R(t)$. We also consider
   that radial turbulent motions could increase the radial growth. Correcting the
   velocity for the radial expansion given by \autoref{eq:rad} as $2c_s^2/v_r$ by
   including the contribution of the density weighted radial turbulence as
   $2(c_s^2+\sigma_r^2)/v_r$ and adding this term to the inflow velocity in the mass
   accretion rate \autoref{eq:mdot}, gives an overestimation for the radial
   growth and a mass accretion rate (dashed-dotted lines in \autoref{fig:dtzero}).
   At about 3.5 Myr the filament reaches the limits of the box and its maximum
   numerically resolved extent. This means that mass cannot be effectively accreted
   and the mass accretion rate goes to zero which leads to a decay in velocity
   dispersion seen in later evolution plots.

   We show a detailed analysis in the next subsections where we present the cases of
   an inflow velocity generating a super- and subsonic internal velocity dispersion.
   This leads to the following conclusion: As long as one includes a symmetry
   breaking perturbation, either numerical or artificial the amount of generated
   turbulence is robust as long as the minimal resolution is high enough.

   \subsection{Trans- and supersonic turbulence}

   In \autoref{fig:evosuper} we plot the evolution of a Mach 5.0 inflow with an initial
   density perturbation. The filament shows every
   indication of turbulent motions rearranging material constantly. The bubbling
   and sloshing in the filament forms temporary ridges and overdensities, the most
   prominent on the central line at the beginning of the simulation. Over time the
   central overdensity weakens as the lack of gravity allows the material to
   spread freely. Nevertheless, the visual impression is that of an observable
   filament albeit the filament broadens to an unrealistic width. The velocity
   dispersion settles to an equilibrium as
   shown in \autoref{fig:sigsuper}. The volume weighted velocity dispersion is in the
   supersonic regime and dominated by the radial velocity dispersion. Interestingly,
   the longitudinal and angular velocity dispersions, both volume and density weighted,
   settle to the same level of about half the sound speed. Furthermore, their values
   are at about 2/3 of the radial and about half the total velocity
   dispersion. This relation between the radial velocity dispersion and the
   other components remains robust and true for all cases of generated velocity
   dispersion, not only in the supersonic case. As the filament reaches the boundary
   of the box earlier than the reference case, the velocity dispersion decays after
   3.5 Myr. Additionally, we show the build-up of the kinetic energy power
   spectrum of the filament in \autoref{fig:superps}. As the filament grows, the
   maximum of the distribution shifts to bigger scales. It always follows closely
   the filament radius, given by the dashed vertical lines. Thus, the radius
   is the scale for the driving mechanism. Included in the plot are the
   predicted scaling relations for the power spectrum. In the supersonic
   regime, numerical studies \citep{kritsuk2007,federrath2010,federrath2013}
   have shown that a pure velocity power spectrum should follow Burgers turbulence
   \citep{burgers1948} with a scaling of $-2$. In the subsonic regime, Kolmogorov's
   theory of incompressible turbulence \citep{kolmogorov1941} predicts a decay to
   smaller scales with a power law of $-5/3$.
   As one can see, the geometrical form of the filament limits the power on large
   scales. Nevertheless, the power spectrum follows the expectation quite well on
   intermediate scales. However, it is impossible to distinguish between supersonic
   and Kolmogorov decay as both of them lie too close together. In \autoref{fig:dpdf}
   we show the volume weighted probability density
   function (PDF). As expected for fully developed, supersonic turbulence it follows a
   log-normal distribution \citep{vazquez1994,padoan1997,passot1998}. The width of the
   distribution is correlated to the Mach number as
   \begin{equation}
     \sigma_s^2 = \ln\left(1+b^2\mathcal{M}^2\right).
     \label{eq:dpdfwidth}
   \end{equation}
   The parameter $b$ depends on the ratio of solenoidal to compressional driving of the
   turbulent motions, varying smoothly from $b=1/3$ for purely solenoidal to $b=1$ for
   purely compressive driving with a value of $b\approx0.4$ for a natural mix of modes
   $F_\mathrm{comp}/(F_\mathrm{sol}+F_\mathrm{comp})=1/3$
   \citep{federrath2008,federrath2010}. All of our simulations with supersonic
   turbulence show about the same value of $b\approx0.55$ but there is also some
   variation over time.

   \subsection{Subsonic turbulence}

\begin{figure*}
      \includegraphics[width=1.75\columnwidth]{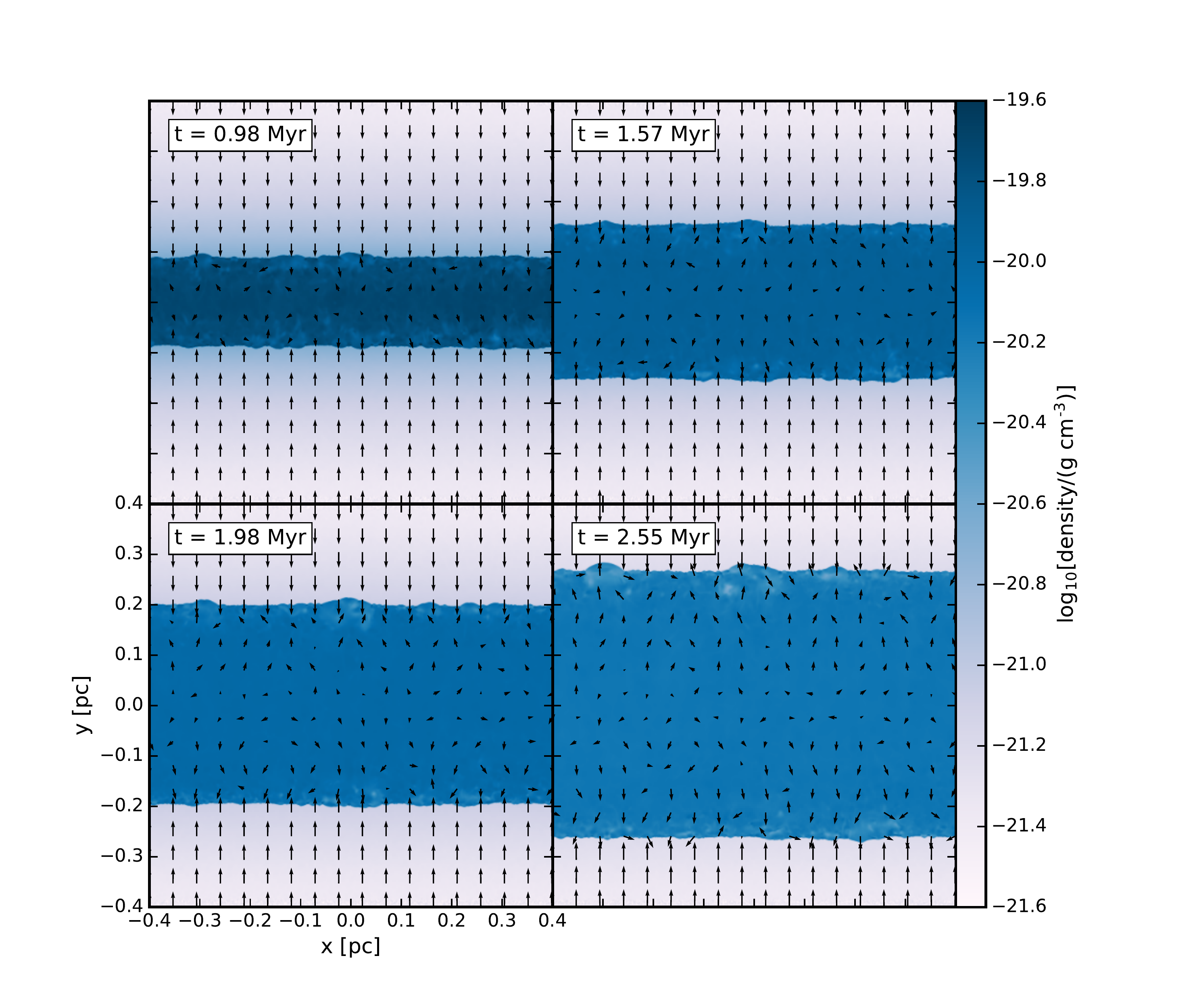}
      \caption{Density cut through the centre of the filament with a Mach 3.0 inflow.
      The velocities are again given by log-scaled arrows. Only the surface of the
      filament is mildly perturbed.}
      \label{fig:evosub}
   \end{figure*}

\begin{figure}
      \includegraphics[width=\columnwidth]{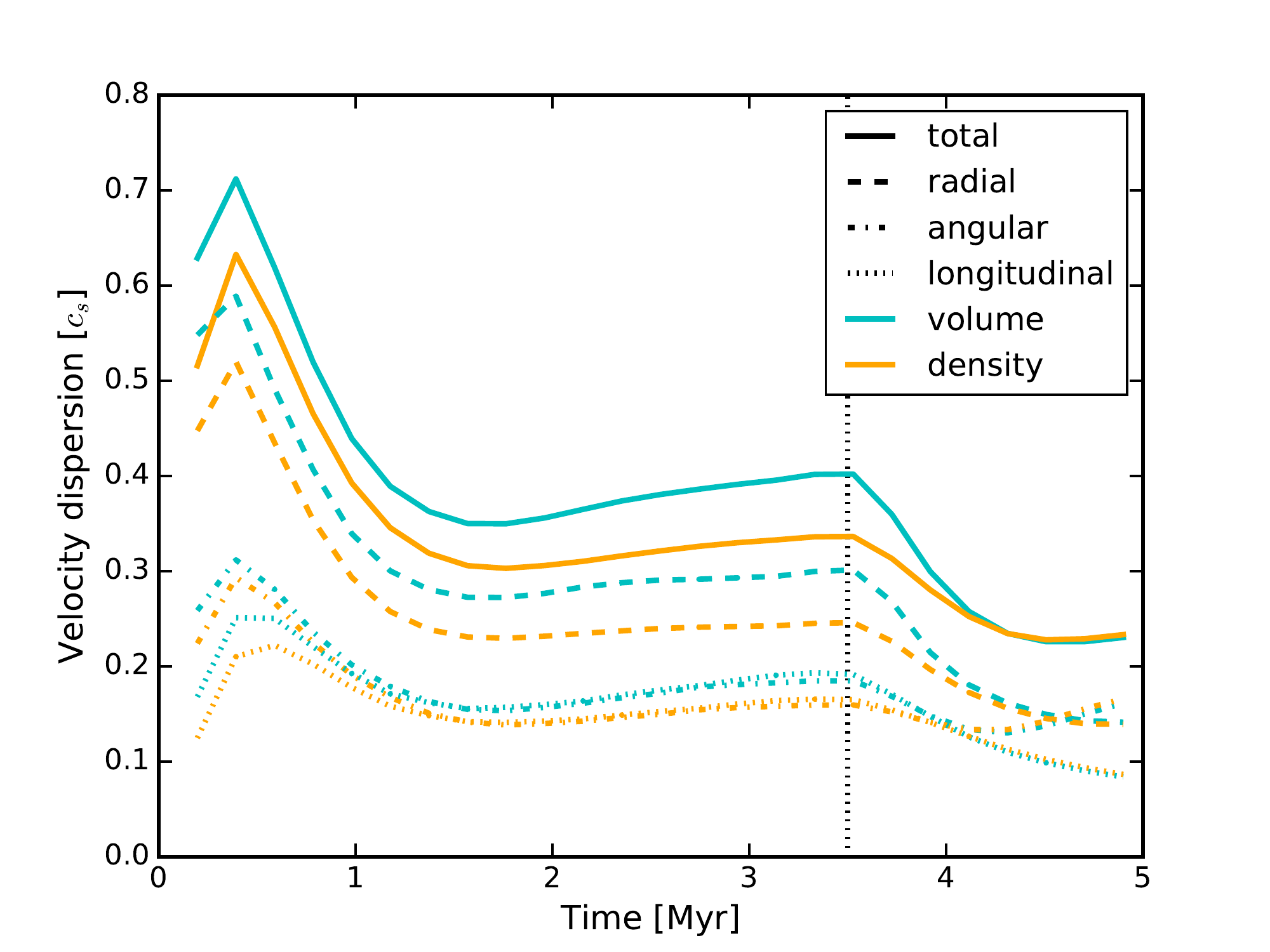}
      \caption{Evolution of the velocity dispersion of a Mach 3.0 inflow. As in the
      case of a Mach 5.0 inflow an equilibrium is established where the velocity
      dispersion is quasi constant albeit there is a slight increase over time.
      After the filament reaches the domain boundary (dotted vertical line) the
      velocity dispersion decays.}
      \label{fig:sigsub}
   \end{figure}

\begin{figure}
     \includegraphics[width=\columnwidth]{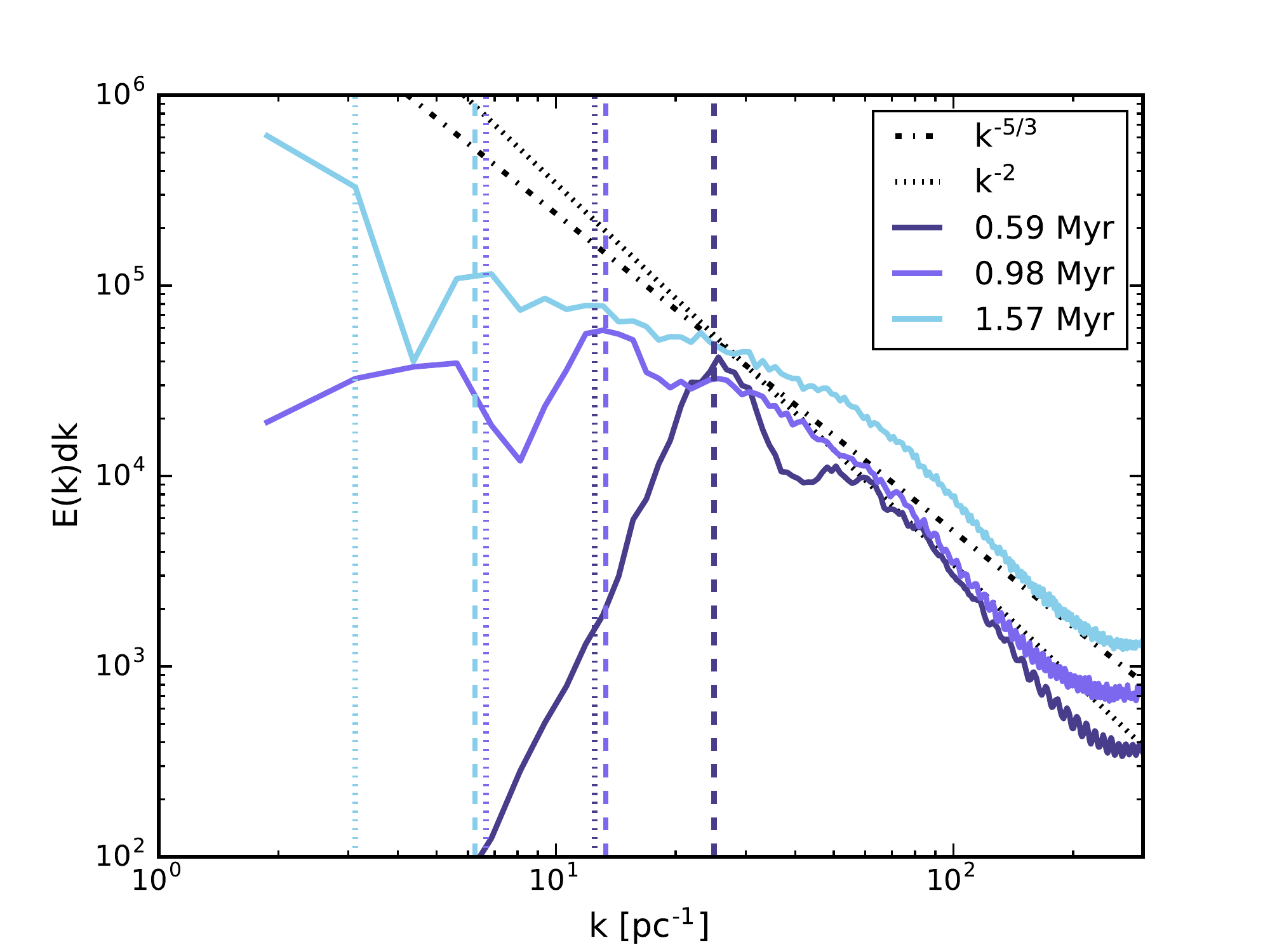}
      \caption{The kinetic energy power spectrum of a filament with Mach 3.0 inflow at
      different times. Similar to the Mach 5.0 case there is a maximum at the
      radial scale length. For late stages, modes along the filament contribute
      significantly to the power spectrum.}
      \label{fig:subps}
   \end{figure}

   In contrast to the turbulent motions of the Mach 5.0 inflow, a slower inflow velocity
   is only capable of generating subsonic turbulent motions despite itself being
   supersonic. In \autoref{fig:evosub} we show the visual impression of a Mach
   3.0 inflow. It is not strong enough to generate substructure inside the filament.
   Only the surface is mildly perturbed. The lack of internal motions can also be seen
   in the velocity dispersion evolution in \autoref{fig:sigsub}. At all times it is
   below the sonic line. Nevertheless, it again reaches an equilibrium after about
   2.0 Myr which decays after reaching the size of the box at about 3.4 Myr. We also
   show the kinetic energy power spectrum for the subsonic turbulent case in
   \autoref{fig:subps}. The power spectrum shows a similar behavior to the
   supersonic case and again the maximum of the cascade, the indicator of the driving
   scale, corresponds to the filament radius. As in the supersonic case, it is not easy
   to distinguish between a Kolmogorov and an $k^{-2}$ cascade but at later times the
   power law seems more similar to the former one which is to be expected for subsonic
   turbulence. In contrast to the supersonic case, we see a strong lack of compressional
   modes in the split-up of the power spectrum. This is also confirmed by
   \autoref{eq:dpdfwidth} where we get a value of $b=1/3$.

   \begin{figure*}
      \includegraphics[width=2.0\columnwidth]{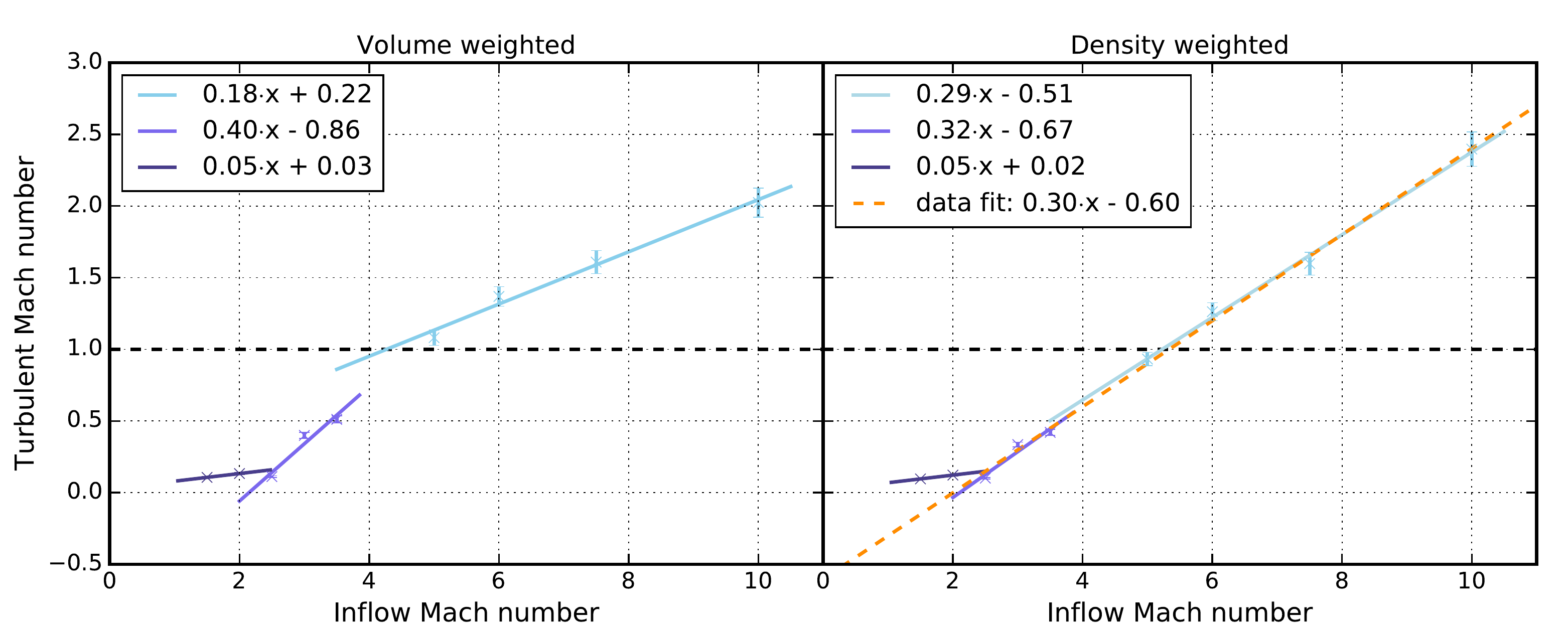}
      \caption{The level of driven velocity dispersion as a function of inflow
      velocity. The volume weighted velocity dispersion is shown on the left
      and the density weighted velocity dispersion on the right. The parameters of
      the fitted lines are given in the legend. The errorbars illustrate the
      inherent variance in the velocity dispersion in different simulations of about
      five per cent. The dashed orange line is fitted over all light and intermediate
      blue values.}
      \label{fig:turb}
   \end{figure*}

   \begin{figure*}
      \includegraphics[width=2.0\columnwidth]{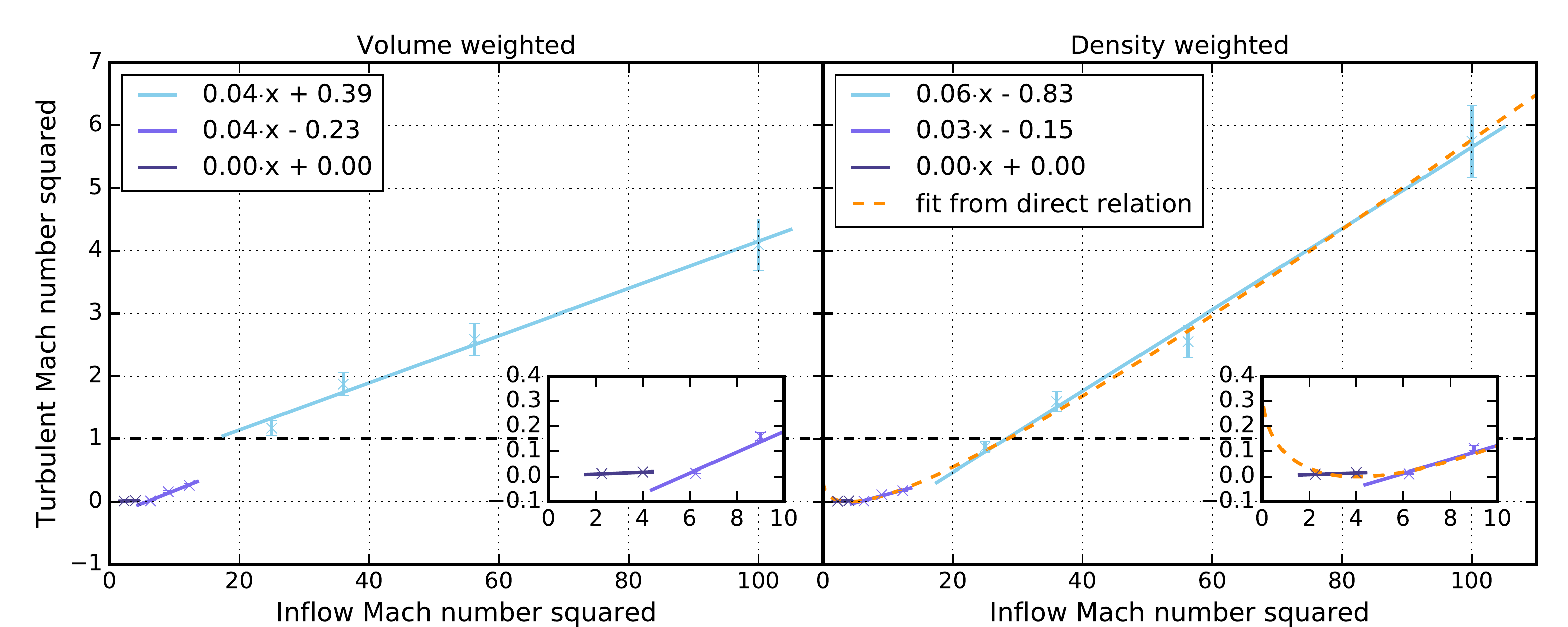}
      \caption{The level of driven turbulent Mach number squared as a
      function of inflow Mach number squared. To calculate the energies one
      has to multiply both terms by half the mass accretion rate. It is a constant
      factor and thus the plot shows also the relation between the respective kinetic
      energies. The volume and density weighted velocity dispersion is shown on the
      left and right respectively. The dashed orange line is the fitted relation from
      \autoref{fig:turb}.}
      \label{fig:turbsqrd}
   \end{figure*}

   \subsection{Dependence on inflow velocity}

   The crucial question of accretion driven turbulence is how much internal velocity
   dispersion in a filament is generated in dependence of the inflow velocity.
   Therefore, we repeat the simulation for inflow velocities of Mach number 1.5, 2.0,
   2.5, 3.0, 3.5, 4.0, 5.0, 6.0, 7.5 and 10.0 for a minimum resolution of
   $512^3$. We show the resulting
   values of the velocity dispersion in \autoref{fig:turb} where we plot them
   against their inflow velocity. The errorbars are given by the inherent variance of
   the velocity dispersion for different initial seeds of the perturbation spectrum
   and is about five per cent of the measured turbulence. In \autoref{fig:turbsqrd} we
   also show the dependence of the Mach number squared on the inflow Mach number
   squared as this is the measure of the turbulent and inflowing energy. Both the
   accreted kinetic energy and the turbulent energy depend on the total accreted mass
   of the filament. Nevertheless, one does not have to account for mass as all the
   accreted material also has to be set into turbulent motions and thus the mass term
   cancels from both energy terms. This is illustrated later in \autoref{eq:efit}.
   In both figures we show the volume weighted velocity dispersion on the left hand side
   and the density weighted velocity dispersion on the right hand side. One can see
   that the measurement method has a major impact on the results. First, we look at the
   volume weighted velocity dispersion in \autoref{fig:turb}. We observe three distinct
   regimes: A high inflow velocity (light blue) leads to supersonic turbulence, an
   intermediate inflow velocity (medium blue) generates subsonic turbulence and
   a low inflow velocity (dark blue) results in nearly no turbulence. All of these
   regimes can be well fitted by linear relationships which do not necessarily go
   through the zero point. The fitting parameters are always given in the legend.
   Striking is the break at the sonic line (dashed horizontal line) going from the
   intermediate subsonic regime to the high velocity supersonic regime. This break
   does not appear in the density weighted velocity dispersion on the right hand side
   of \autoref{fig:turb}. Now the high and medium inflow velocity values follow one
   and the same relation. Thus, we now can fit the data by one single linear
   relationship that connects both regimes and that is shown by the dashed orange line
   with the parameters:
   \begin{equation}
     \sigma = 0.30v_r - 0.60 c_s.
     \label{eq:fit}
   \end{equation}
   The fact that we can fit a single line from the high supersonic end down to low
   subsonic turbulent velocities shows that there is no difference in the physics at
   work in both regimes. The break in the volume weighted relation also occurs for the
   squared volume weighted velocity in \autoref{fig:turbsqrd}. Although the slope
   stays roughly the same there is a gap between the supersonic and the subsonic regime.
   Contrarily, the density weighted velocity dispersion squared shows a smooth
   transition from nearly no turbulence to a constant slope. The density weighted
   velocity dispersion is also a measurement of the kinetic energy in the turbulence
   of the filament, a fact that we have confirmed by calculating the kinetic energy.
   The break in the volume weighted velocity dispersion results from the fact
   that the nature of the velocity and density distribution changes from
   the subsonic to the supersonic regime. Supersonic turbulence is shock dominated and
   due to compression most of the kinetic energy is in high densities. The information
   about the high densities is lost if one only measures the volume weighted velocity
   dispersion.

   In the right panel of \autoref{fig:turbsqrd} we also overplot \autoref{eq:fit}.
   Despite being a linear correlation between velocity dispersion and infall speed it
   fits the numerical results remarkably well and deviates only slightly for very low
   Mach numbers. We also attempt to fit a parabola to all of the density weighted data
   but are not able to get a good match. Therefore, we conclude that the linear
   relation of \autoref{eq:fit} provides the best analytical description of the
   underlying physics. While the data points could also be fitted reasonable
   well with an offset power law of 1/3rd, they do not follow the simple prediction of
   \citet{heitsch2013}, given by \autoref{eq:eequ}. We discuss the implication of this
   in \autoref{sec:discussion}.

   We repeat a subset of the simulations for a higher temperature of 40 and
   100 K and get similar results, all lying on the analytical relation given by
   \autoref{eq:fit}. Thus our results are independent of the temperature. We also
   repeat a subset of the simulations with an additional longitudinal velocity
   component in the inflow. The result in velocity dispersion is unchanged and the only
   difference is that the whole filament now begins to move in x-direction with
   constant velocity.

\section{Pressure equilibrium}
\label{sec:pequilibrium}

   \begin{figure}
      \includegraphics[width=\columnwidth]{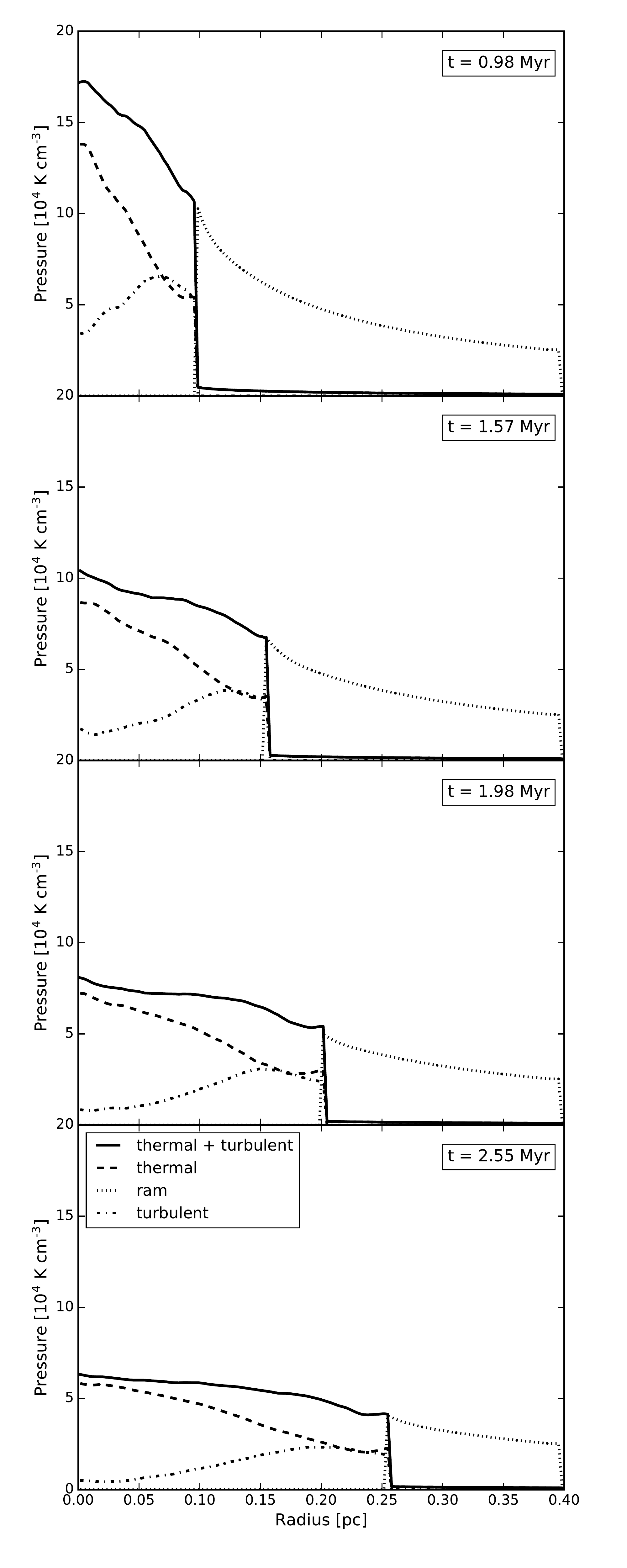}
      \caption{Radial pressure profile of the Mach 5.0 case with an initial density
      perturbation for different times. Inside the filament a pressure equilibrium is
      established through the combination of thermal and turbulent pressure.}
      \label{fig:prad}
   \end{figure}

   In order to see the impact of turbulence on the pressure we analyse its components
   as function of radius and time. In \autoref{fig:prad} we show the different
   contributions to the pressure calculated in radial bins for the Mach 5.0 case with
   an initial density perturbation. We calculate the velocity dispersion, the
   average density and average thermal pressure in each slice along the filament and
   in radial bins which are chosen to be four cells wide. We use these values to
   calculate the respective pressure terms and take the mean along the filament in
   order to determine an average value for the whole filament. There is an overlap
   region where filament gas and environment gas mix due to the filament not being
   completely straight and round. Therefore, we only use the filament gas to
   calculate the pressure components in the overlap region. We determine the turbulent
   pressure component with the density weighted velocity dispersion. Thus, the
   turbulent pressure is given by
   \begin{equation}
     P_{\mathrm{turb}} = \left<\rho\right>\sigma^2,
   \end{equation}
   where the braket notation represents the expectation value. The average ram pressure
   is calculated by the average density and radial velocity as:
   \begin{equation}
     P_{\mathrm{ram}} = \left<\rho\right> \left<v_r\right>^2
   \end{equation}

   One can see in \autoref{fig:prad} that outside of the filament the turbulent
   pressure (dashed-dotted line) is zero and the thermal pressure (dashed line) is
   negligible. As material streams towards the filament the ram-pressure (dotted
   line) increases as the density increases. When the accretion flow reaches the
   filament the ram-pressure breaks down and some of the energy is converted into
   turbulent motions giving rise to a turbulent pressure component. In the overlap
   region of filament and environment gas we see both a contribution of turbulent
   pressure and ram pressure due to our split-up of components. As one can see the
   average pressure inside the filament given by the average thermal pressure
   together with the average turbulent pressure (solid line) has nearly the same
   value as the average ram pressure. The filament is compressed and restrained by
   the accretion flow. Before the filament has settled there is quite an overpressure
   in the centre of the filament. As the filament grows outward, it adjusts to the
   outside ram-pressure and a constant pressure inside the filament is established.
   Note that there is a gradient in the pressure components inside the filament
   where the thermal pressure decreases and the turbulent pressure increases towards
   the edge of the filament. Most of the turbulent motions are created at the boundary
   and propagate inwards. This also explains why there is no turbulent component
   supporting the filament radially as the turbulence acts in a localized region.

\section{Observable velocity dispersion}
\label{sec:observations}

   \begin{figure*}
     \includegraphics[width=2.0\columnwidth]{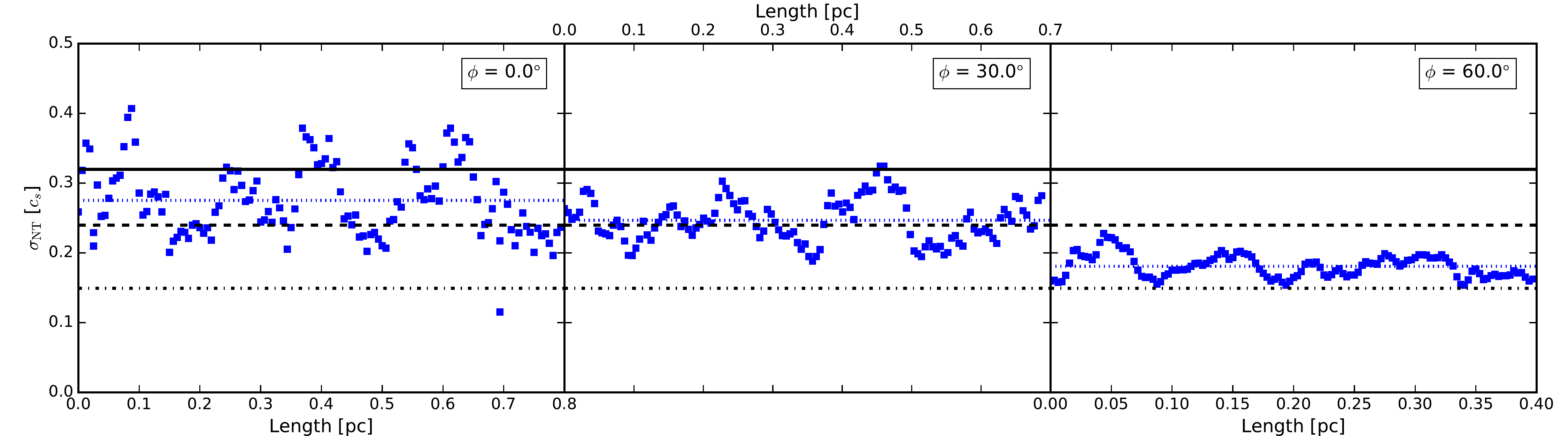}
     \caption{The measured non-thermal linewidth along the central axis of the filament
     for the case with a Mach 3.0 inflow for different inclinations. Data
     points are given by the blue squares and their mean by the blue dotted line. The
     solid, dashed and dashed-dotted black lines show the total, the radial and
     the longitudinal density weighted velocity dispersion respectively.}
     \label{fig:vlossub}
   \end{figure*}

   \begin{figure*}
     \includegraphics[width=2.0\columnwidth]{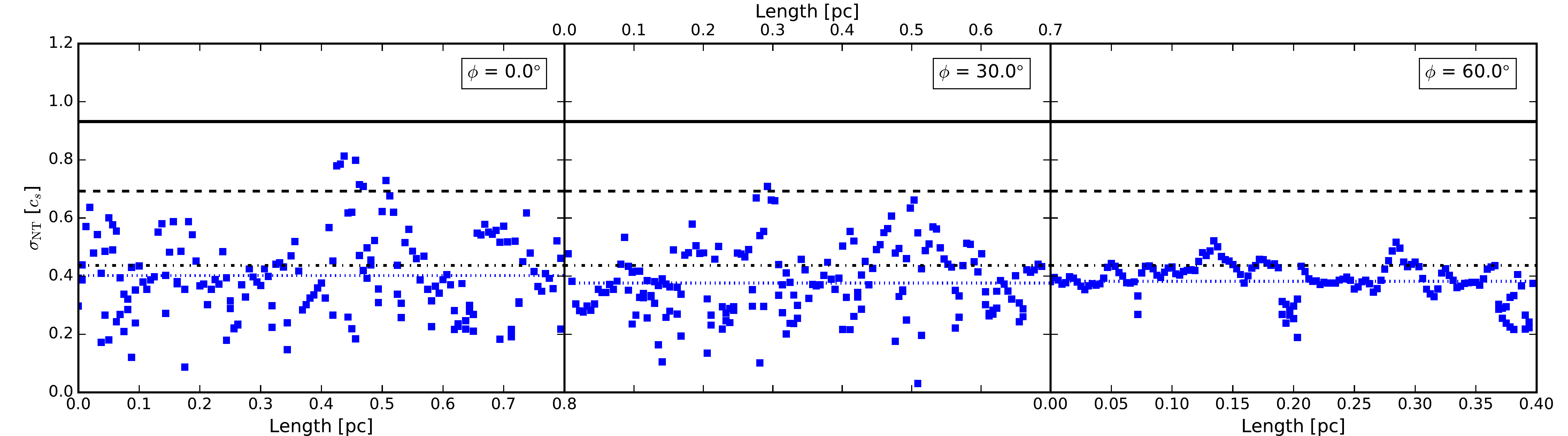}
     \caption{The same as in \autoref{fig:vlossub} but for an inflow of Mach 5.0.}
     \label{fig:vlossup}
   \end{figure*}

   \begin{figure*}
     \includegraphics[width=2.0\columnwidth]{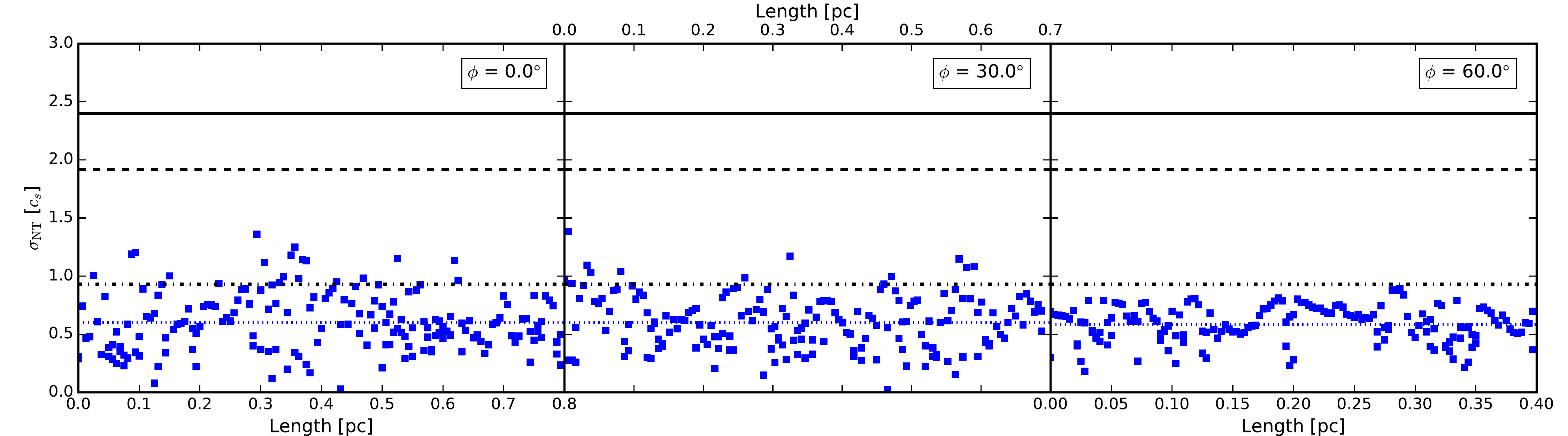}
     \caption{The same as in \autoref{fig:vlossub} but for an inflow of Mach 10.0.}
     \label{fig:vlos10}
   \end{figure*}

   In order to compare our simulations to observations we derive the line-of-sight
   velocity by sending a ray through the computational domain. We vary the inclination
   of the filament and treat every crossed volume element with its respective density
   as a discrete emitter of a line-of-sight velocity value. These are converted to
   density weighted Gaussian line profiles with a dispersion width corresponding to the
   thermal linewidth of \CO\ at 10 K with a given value of $\sigma = 0.0526$ \kms. We bin
   the resulting line profiles into histograms with a bin width of $0.05$ \kms\ to get
   a complete line emission for each observed spatial pixel. We then measure the
   velocity centroids and the linewidths as an observer would do by fitting Gaussians to
   the line. As several line profiles show multiple components in velocity space, we
   fit multiple Gaussians to a single line. In contrast, fitting only one Gaussian to a
   line overestimates the velocity dispersion as multiple components are grouped
   together to form one broad line. To prevent false detections we discard peaks which
   are smaller than a quarter of the maximum line profile and closer together than
   1.5 times their respective width. To obtain the non-thermal velocity dispersion we
   have to subtract the contribution of the thermal gas motions of the full width half
   maximum
   (FWHM) of the line as e.g. \citet{myers1983}:
   \begin{equation}
     \sigma_\mathrm{NT} = \sqrt{\frac{\text{FWHM}^2}{8\ln 2} - \frac{k_B T}{m}}
   \end{equation}
   We show the non-thermal linewidth along a filament through its centre
   for the case of turbulence forming due to inflows of Mach 3.0, 5.0 and 10.0 in
   \autoref{fig:vlossub}, \autoref{fig:vlossup} and \autoref{fig:vlos10}, all of which
   include an initial perturbation.
   At a first glance it is important to note the striking similarity to real
   observations \citep{hacar2011,hacar2013,tafalla2015,hacar2016}. As in the actual
   measurements we see an inherent spatial variation due to the non-homogeneous nature
   of the turbulence. The variation is also not stationary but changes with time, mimicking
   the sloshing motions of the total filament. However, the mean of the data
   distribution stays relatively stable, only varying by about 10\%. Comparing
   the mean to the analytical value of the velocity dispersion it becomes clear that
   an observer cannot see the full picture of the internal motions of the filament.
   The observer will preferentially see the density weighted radial velocity
   dispersion (dashed line) as filaments are more likely to be detected for small
   inclinations. Thus, the line-of-sight is parallel to the radial motions. While the
   subsonic case shows mean values which lie below any velocity dispersion typically
   observed, it is also the only one where the mean of the observed distribution lies
   above the expected value of the radial velocity dispersion. Going to higher inflow
   speeds, the mean of the observed distribution generally lies below the expected
   value. The situation becomes even worse for cases of high turbulence as shown in
   \autoref{fig:vlos10} where the discrepancy can be even as large as a factor of two.
   Moreover, the discrepancy between the observed mean and the total velocity
   dispersion is even larger. Despite having a supersonic total density weighted
   velocity dispersion, most of the data points of the Mach 10 inflow show a subsonic
   or at best a transsonic non-thermal line width. This means that the interpretation
   of the inherent motions in an observed filament can be severely flawed. A
   filament dominated by supersonic motions will be classified as only having
   subsonic motions. With the addition of gravity the problem will become
   even worse as there is an internal density gradient enhancing the signal of the
   central region. This leads to an overall thinner line as the outer region,
   which contains a high radial velocity component, is neglected and thus the
   measured line width becomes even smaller. There is not even a constant correction
   factor as for example assumed for isotropic turbulence of $\sqrt{3}$. The
   measured mean is relatively constant for different inclinations, going from zero to
   sixty degrees, but if one does not split the individual components in the line its
   broadness is overestimated the more the filament is seen from the side. The larger
   the inclination, the more is the filament dominated by longitudinal motions which
   give only one thin component. Therefore, the observable mean decreases and finally
   approaches the longitudinal velocity dispersion. This can be seen to some degree in
   the subsonic case. If one fits one component to the line-of-sight this effect is
   stronger and can also be observed in the cases of higher inflow velocity. It is
   even possible for the mean to be below the longitudinal velocity dispersion as a
   single line-of-sight does not contain the complete information on the total
   line-of-sight velocity dispersion of the whole filament. Striking is also the fact that the
   variance of the distribution decreases for a higher inclination. This behavior could
   potentially be used to determine the inclination of a filament. We will study the
   effect of the inclination angle on the measurements of the turbulent velocity and
   its variance in a subsequent paper.

\section{Discussion and conclusions}
\label{sec:discussion}

   We have presented a numerical study on accretion driven turbulence in filaments.
   The focus of this paper was to analyse the dependence of the velocity dispersion
   on the inflow velocity. We deliberately neglected the effects of gravity in order
   to have a constant inflow velocity and to be able to follow the driving of
   turbulence long enough without the radial collapse of the filament or its
   condensation into collapsing clumps. The formation of turbulent filament, including
   self-gravity will be discussed in a subsequent paper. We find that
   there is a linear dependency of the density weighted velocity dispersion
   on the inflow velocity. Below Mach 2.0 the relationship flattens as shown in the
   right panels of \autoref{fig:turb} and \autoref{fig:turbsqrd}. Why there
   is a break in the relation is not completely clear. One possibility is that the low
   inflow velocity region is an artifact of numerical noise. In this case it must be
   dominated by perturbations on small scales. Therefore, we smooth our data to
   remove the small scale signal and recalculate the velocity dispersion. The result
   shows that the velocity dispersion is unchanged. Therefore, the measured velocity
   dispersion is in the large scale structure. Even for a high degree of smoothing the
   velocity dispersion remains the same. Thus, there must be a physical reason for the
   break in the slope. An explanation could be that filaments have a bottom level of
   minimal velocity dispersion comparable to a basic eigenmode which is exited by the
   constant inflow. In the subsonic regime sound waves are supposed to be very
   inefficient in dissipating energy and indeed if one sets up a filament with a random
   velocity perturbation there is a level of velocity dispersion below which the
   kinetic energy does not decay. Its dependence on form and density of the filament
   is out of the scope of this paper and will be discussed elsewhere.

   We now want to focus on the form of the linear fit. As shown in \autoref{fig:turb} we
   measure a linear relationship of
   \begin{equation}
     \sigma = 0.30v_r-0.60c_s
   \end{equation}
   It is possible to show that this is equivalent to an energy balance. In order to
   get a kinetic energy we take the square of the equation and multiply it with half
   of the mass accretion rate which gives
   \begin{equation}
     \half \dot{M} \sigma^2 = 0.09\cdot\half \dot{M} v_r^2 - 0.18\dot{M}c_sv_r + 0.18\dot{M}c_s^2
     \label{eq:efit}
   \end{equation}
   with the restriction that
   \begin{equation}
     v_r \ge 2.0c_s.
   \end{equation}
   For a constant velocity dispersion the term on the left hand side is the change
   in turbulent energy and the first term on the right hand side the rate of accreted
   kinetic energy. To show the meaning of the remaining terms we use the equation of change
   in turbulent energy given by
   \begin{equation}
      \dot{E}_t = \alpha \dot{E}_a - \dot{E}_d
      \label{eq:model2}
   \end{equation}
   where $\dot{E}_t$ is the change in turbulent energy, $\dot{E}_a$ is the kinetic
   energy accretion rate and $\dot{E}_{d}$ is the energy dissipation rate. In the case
   of a constant velocity dispersion, the change in turbulent energy is due to the
   change in total mass of the filament. The same is true for a constant accretion
   velocity where the change in total accreted kinetic energy is due to the change in
   the total accreted mass. We also consider that energy is lost in the isothermal
   accretion shock and is radiated away. As the fraction of the lost energy should be
   roughly constant per time we introduce a constant efficiency $\alpha$ which has a value
   between zero and one. Comparing \autoref{eq:model2} to \autoref{eq:efit} we determine
   the efficiency to be:
   \begin{equation}
     \alpha = 0.09.
   \end{equation}
   This means that the energy lost to dissipation is given by the remaining terms
   \begin{equation}
     \dot{E}_d = 0.18\dot{M}c_sv_r - 0.18\dot{M}c_s^2 = 0.18\dot{M} \left( c_sv_r - c_s^2 \right)
     \label{eq:modeldiss}
   \end{equation}
   A possible explanation for both terms is the assumption that turbulence is mainly
   dissipated in an inner region of the filament where radial waves travelling inwards
   interact. We assume that this region is proportional to the total radius of the filament:
   \begin{equation}
     R_i = b R(t) = b \frac{2c_s^2t}{v_r}
   \end{equation}
   The proportionality $b$ has to be constant for a single run but can vary with accretion
   velocity. The mass contained in the inner region is then the density given by the shock
   times the volume:
   \begin{equation}
     M_i = \rho(R) \pi R_i^2 L = \rho_0 \mathcal{M}^2 R_0 \pi b^2 R L = b^2 M(R)
   \end{equation}
   This means that the mass contained in a region proportional to the radius is also
   proportional to the total mass. Thus, if we use the definition of the energy dissipation
   rate given by \autoref{eq:ediss}, assume that the scale relevant for the crossing time
   is the radius of the inner region and assume that the speed for the crossing time is the sound
   speed we get:
   \begin{equation}
     \dot{E}_d = \half \frac{M_i c_s^3}{R_i} = \frac{b}{4} \frac{M c_s^3 v_r}{c_s^2t}
               = \frac{b}{4} \dot{M} c_s v_r
   \end{equation}
   As the inner region should be larger for greater inflow Mach number $\mathcal{M}$ we assume that:
   \begin{equation}
     b = a\left(1-1/\mathcal{M}\right)
   \end{equation}
   where $a$ is a constant. Thus the energy decays as:
   \begin{equation}
     \dot{E}_d = \frac{a}{4} \left(1-c_s/v_r\right) \dot{M}c_s v_r =
                 \frac{a}{4} \dot{M}(c_s v_r - c_s^2)
   \end{equation}
   If we compare our result to \autoref{eq:modeldiss} we find this fits both terms
   with $a=0.72$.\\

   \noindent Our results can be summarized as follows:
   \begin{enumerate}
     \item The accretion of material leads to non-isotropic, sub- and supersonic
           turbulence depending on the energy accretion rate as long as the symmetry
           is broken. The efficiency of transferred energy is independent of the
           accretion rate and equal to 9\%.
     \item The amount of turbulence generated is independent of the perturbation
           strength of the initial density field and independent of an
           additional velocity component parallel to the filament.
     \item The turbulence reaches an equilibrium level which scales linearly with the
           accretion rate.
     \item The filament radius grows linearly with time and is the driving
           scale of the turbulence.
     \item The density weighted velocity dispersion contributes to the pressure
           equilibrium inside the filament.
     \item In most situations, an observer cannot measure the true velocity
           dispersion and will even misinterpret the level of turbulence to be subsonic
           while the intrinsic velocity dispersion could still be supersonic.
   \end{enumerate}

\section*{Acknowledgements}

   We thank the whole CAST group for helpful comments and discussions.
   AB, MG and SH are supported by the priority programme 1573
   "Physics of the Interstellar Medium" of the German
   Science Foundation and the Cluster of Excellence
   "Origin and Structure of the Universe". The simulations were run using
   resources of the Leibniz Rechenzentrum (LRZ, Munich; linux cluster CoolMUC2).
   We would also like to thank the referee whose input lead to an improvement
   of the results of the paper.



\bibliographystyle{mnras}
\bibliography{Turb}







\bsp	
\label{lastpage}
\end{document}